\documentclass[aip, amsmath,amssymb,
preprint,%
nofootinbib, tightenlines,cha,
]{revtex4-1}

\usepackage[pdftex,bookmarks, colorlinks, citecolor =blue, breaklinks]{hyperref}

\usepackage{graphicx}
\usepackage{multirow}
\usepackage{dcolumn}

\usepackage[multidot]{grffile} 
\usepackage{bm}

\usepackage[utf8]{inputenc}
\usepackage[T1]{fontenc}
\usepackage{mathptmx}
\usepackage{float}
\usepackage[subrefformat=parens,labelformat=parens]{subfig}
\usepackage{xcolor}
\usepackage{xspace}

\newcommand{\bN}{{\mathbb{ N}}}

\newcommand{\bR}{{\mathbb{ R}}}

\newcommand{\tC}{\tau_C}
\newcommand{\tAMI}{\tau_{I}}

\begin{document}

\preprint{Currently in review at Chaos: An Interdisciplinary Journal of Nonlinear Science}

\title{Using Curvature to Select the Time Lag for Delay Reconstruction}
\author{Varad Deshmukh}
\affiliation{Department of Computer Science, University of Colorado, Boulder, CO, USA}
\author{Elizabeth Bradley}
\affiliation{Department of Computer Science, University of Colorado, Boulder, CO, USA}
\affiliation{Santa Fe Institute, Santa Fe, NM, USA}
\author{Joshua Garland}
\affiliation{Santa Fe Institute, Santa Fe, NM, USA}
\author{James D. Meiss}
\affiliation{Department of Applied Mathematics, University of Colorado, Boulder, CO, USA}

\date{\today}

\begin{abstract}

We propose a curvature-based approach for choosing good values for the
time-delay parameter $\tau$ in delay reconstructions.  The idea is
based on the effects of the delay on the geometry of the
reconstructions.  If the delay is chosen too small, the reconstructed
dynamics are flattened along the main diagonal of the embedding space;
too-large delays, on the other hand, can overfold the dynamics.
Calculating the curvature of a two-dimensional delay reconstruction is
an effective way to identify these extremes and to find a middle
ground between them: both the sharp reversals at the ends of an
insufficiently unfolded reconstruction and the folds in an overfolded
one create spikes in the curvature.  We operationalize this
observation by computing the mean 
over the Menger curvature
of 2D reconstructions for different time delays.  We show that the
mean of these values gives an effective heuristic for choosing the
time delay.  In addition, we show that this curvature-based heuristic
is useful even in cases where the customary approach, which uses average mutual
information, fails---{\sl e.g.}, noisy or filtered data.

\vspace*{1ex}
\noindent
\end{abstract}

\maketitle

\begin{quotation}

Delay-coordinate reconstruction, the foundation of nonlinear
time-series analysis, involves two free parameters: the embedding
dimension $m$ and the delay $\tau$.  A number of heuristic methods are
available for choosing good values for these parameters: notably, the
false near neighbor method of Kennel {\sl et al.} for $m$ and the
average mutual information (AMI) of Fraser \& Swinney for $\tau$.  The
AMI approach selects a $\tau$ that attempts to produce independent
coordinates in the reconstructed trajectories.
Taking a geometric view of this problem, we develop a curvature-based
method for this task.  By computing statistics on the curvature of a
2D reconstruction, we can identify a delay that unfolds the dynamics
without introducing overfolding: {\sl i.e.}, between the extremes that
can cause the embedding to not be a faithful representation of the
full state-space dynamics.  As in AMI, this involves identifying the
first minimum in a plot of the statistic versus $\tau$---something
that is sometimes difficult with AMI because the minima can be shallow
or even nonexistent.  Using a suite of examples, we demonstrate that
the minima in curvature-based statistics are effective in producing
embeddings whose correlation dimensions match those of the true
dynamics.  The curvature heuristic is quite robust in the face of data
issues like noise, smoothing, and shorter samples of the dynamics, and
its minima are generally more distinct, which makes the choice easier.
\end{quotation}

\section{Introduction}\label{sec:Intro}

\begin{figure*}
	\subfloat[]{
	\includegraphics[width=0.32\linewidth]{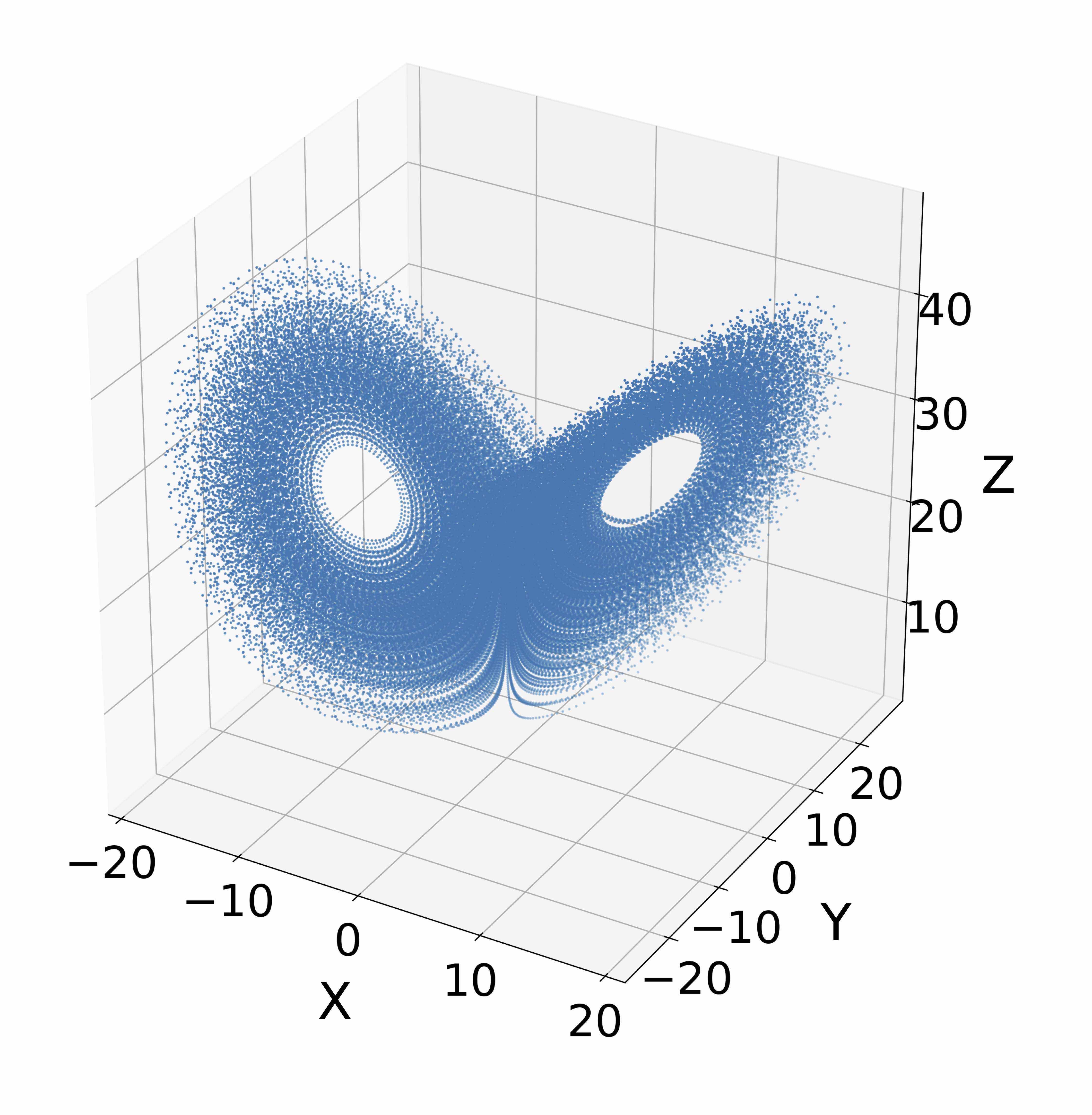}
	}
	\subfloat[]{
	\includegraphics[width=0.32\linewidth]{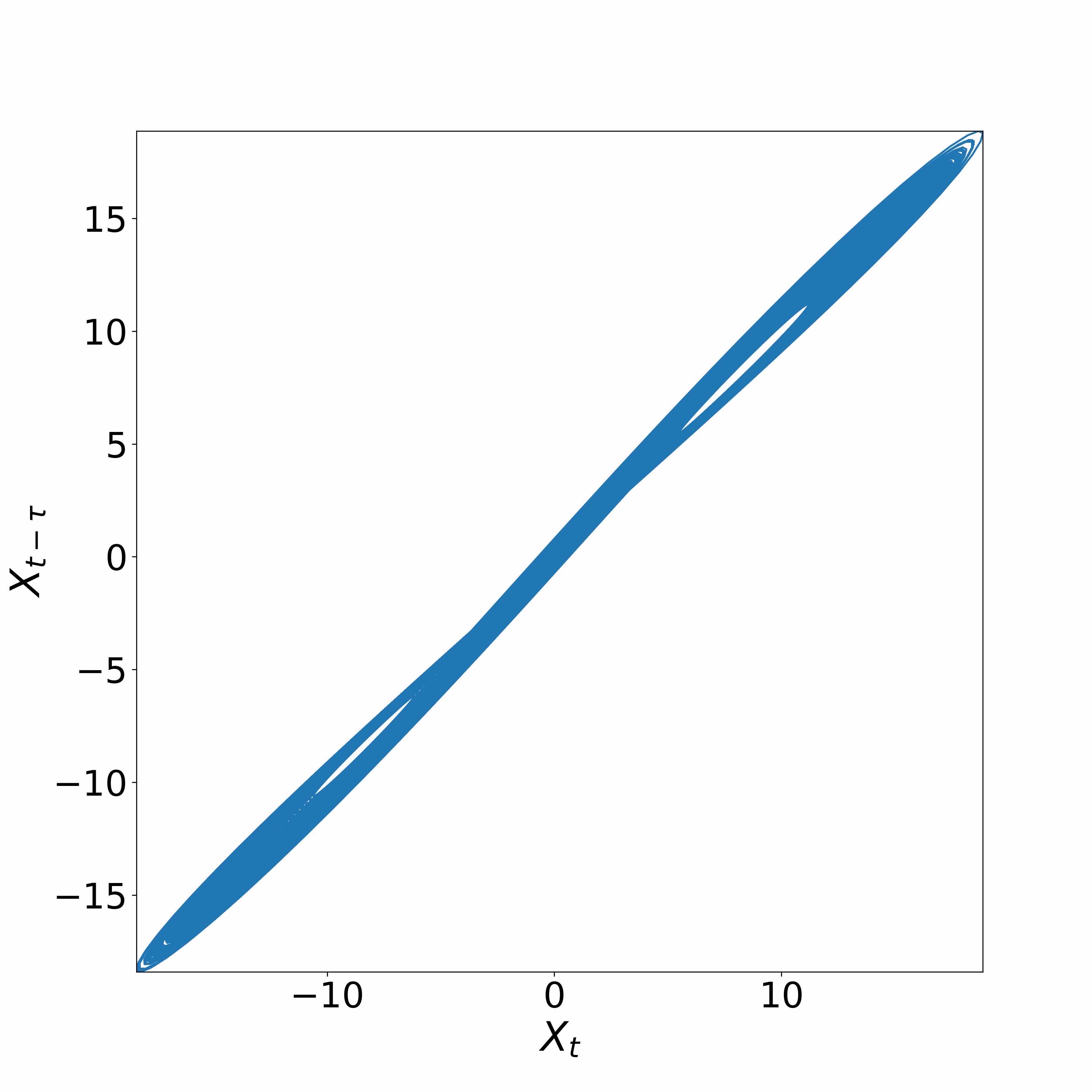}}
	\subfloat[]{
	\includegraphics[width=0.32\linewidth]{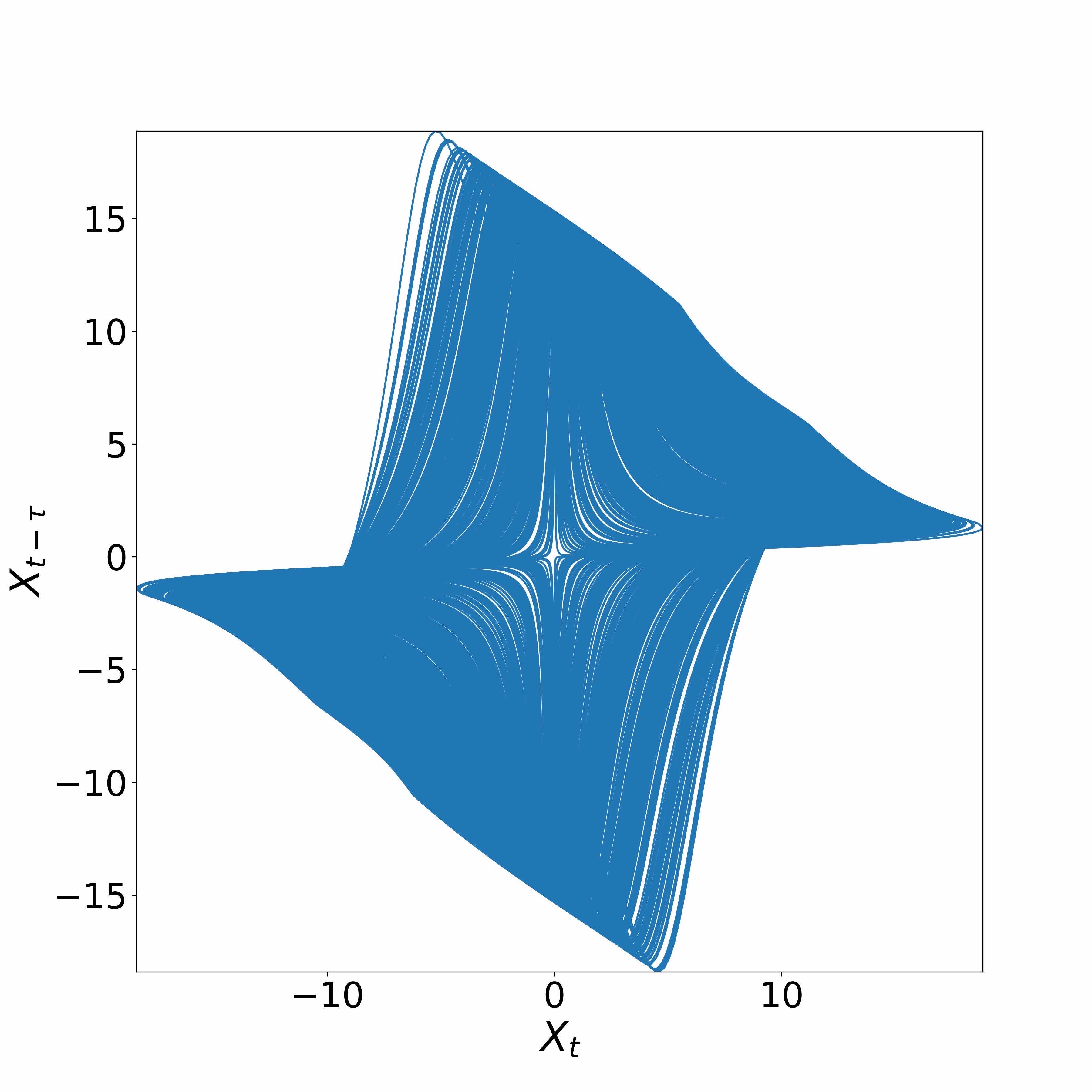}}
	\caption{The Lorenz attractor: (a) full state-space trajectory; 2D delay
          reconstructions from the $x(t)$ time series with (b)
          $\tau=1$ and (c) $\tau=30$.}
	\label{fig:lorenz}
\end{figure*}

Delay-coordinate embedding, or the method of delays, is a
well-established technique for dynamical reconstruction of time-series
data.\cite{takens,sauer91,packard80} This method, which generates a
reconstruction by plotting scalar time-series data against delayed
versions of itself on orthogonal axes, involves two free
parameters---a time delay, $\tau$, and an embedding dimension, $m$.
Examples of reconstructions for the classic Lorenz attractor are shown
in Fig.~\ref{fig:lorenz}.  Though the embedding theorems offer some
theoretical guidance regarding the selection of $\tau$ and $m$, one
must fall back upon heuristics for choosing their values when faced
with finite-precision data from an unknown system.

Over the past decades, the nonlinear dynamics community has devoted
significant effort to developing effective methods for estimating $m$
and $\tau$.  This paper offers a contribution to that arsenal: a new,
geometry-based method for determining $\tau$.  The challenge is this:
if the delay is too small, the reconstructed dynamics are flattened
along the main diagonal of the embedding space; see, {\sl e.g.},
Fig.~\ref{fig:lorenz}(b).  Delays that are too large, on the other
hand, can overfold the dynamics, as in Fig.~\ref{fig:lorenz}(c).  A
common approach to solving this problem, due to Fraser \& Swinney,
\cite{fraser-swinney} seeks a $\tau$ that mazimizes independence
between $\tau$-separated points in the time series, thereby separating
the trajectories. As an alternative, we propose to use the {\sl
  curvature} of trajectories to find a $\tau$ that effectively unfolds
the dynamics while avoiding overfolding.  The idea is based upon the
observation that both sharp reversals like those in
Fig.~\ref{fig:lorenz}(b) and overfolds, as in in
Fig.~\ref{fig:lorenz}(c), can create regions of large curvature in a
2D projection of the reconstructed dynamics.

\begin{figure}
	\includegraphics[width=3in]{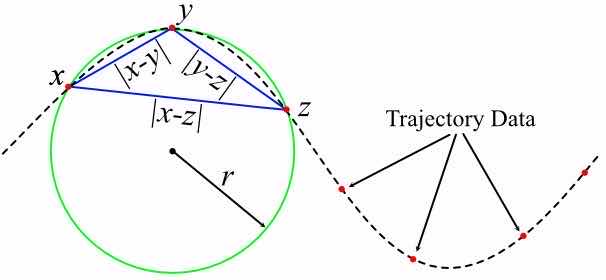}
	\caption{The local curvature of a triplet of successive points
          selected from a trajectory of a dynamical system.}
	\label{fig:menger}
\end{figure}

We operationalize this observation by computing various statistics
over the local curvature of 2D reconstructions using a discrete
curvature due to Menger: the curvature, $c$, of three non-collinear
points---say $x$, $y$, and $z$---is the inverse of the radius of the
unique circumcircle through these points.\cite{Blumenthal70} It can
be seen that
\begin{equation}\label{eq:menger}
c(x,y,z) = \frac{1}{r} = \frac{4A}{|x-y||y-z||z-x|},
\end{equation}
where $A$ is the area of the triangle $xyz$.  In our case, as
illustrated in Fig.~\ref{fig:menger}, $x$, $y$, and $z$ are three
successive points along a reconstructed trajectory of a dynamical
system, and we define
\begin{equation}\label{eq:localC}
	M_j(\tau) = c(\vec{x}_{j-1},\vec{x}_j,\vec{x}_{j+1}) .
\end{equation}
Of course, when the trajectory is relatively straight, the local
curvature will be small; if it has a sharp turn, the local curvature
will be large.

\begin{figure*}[htb!]
	\includegraphics[width=0.8\linewidth]{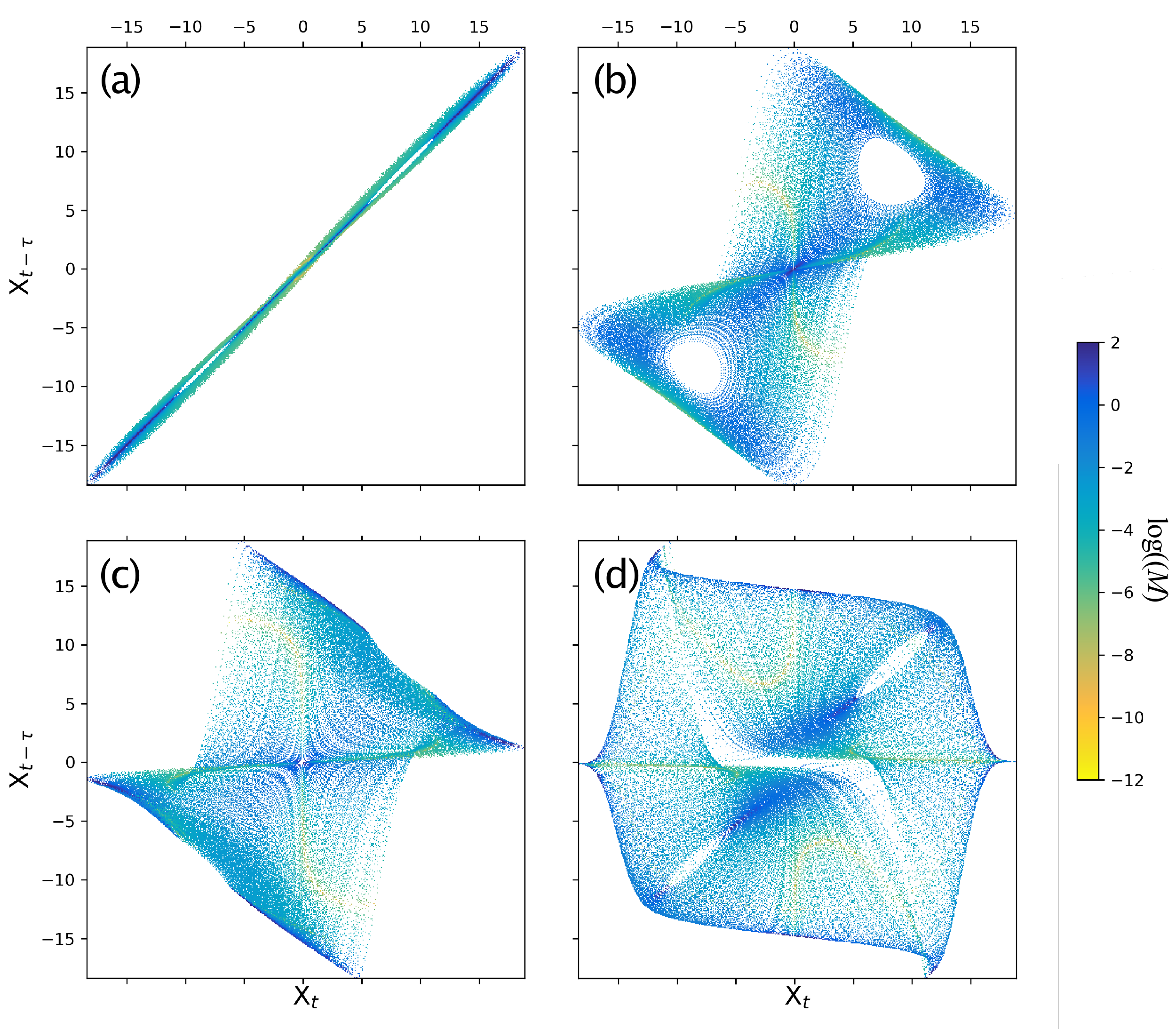}
\caption{Curvature heat maps of 2D delay reconstructions of the Lorenz
  attractor of Fig.~\ref{fig:lorenz}: (a) $\tau = 1$, (b) $\tau = 18$,
  (c) $\tau = 30$, and (d) $\tau = 60$. The color represents the local value of the
  log of the curvature, \eqref{eq:localC}.  Panels (a) and
  (c) correspond to panels (b) and (c) of Fig.~\ref{fig:lorenz},
  respectively.} \label{fig:lorenz_curvature_heatmaps}
\end{figure*}

The variations of the curvature along a trajectory can be visualized
by computing a local average.  For example, 2D reconstructions of data
from the Lorenz attractor are shown in
Fig.~\ref{fig:lorenz_curvature_heatmaps} for four different values of
the time delay $\tau$.  In each case, the color represents the local
average of \eqref{eq:menger} over trajectory points that fall in a
grid cell, using a uniform $500 \times 500$ grid.  These heat maps
bring out the effects of $\tau$ upon the local curvature.  When $\tau$
is small, the reconstructed trajectory has sharp turns at its local
maxima and minima, which manifest as dark blue regions in
Fig.~\ref{fig:lorenz_curvature_heatmaps}(a).  Spikes in the curvature
can also arise from overfolding, when the time delay is large, as in
Fig.~\ref{fig:lorenz_curvature_heatmaps}(c).  We seek a middle ground
between these two extremes by choosing a time delay that minimizes the
average curvature, $\overline{M}$, along a 2D reconstructed
trajectory; as we will see, this corresponds to
Fig.~\ref{fig:lorenz_curvature_heatmaps}(b).
Specifically, the proposed heuristic is that $\tau$ be chosen to give
the first minimum of $\overline{M}(\tau)$.

In \S\ref{sec:Results}, we evaluate the effectiveness of this
curvature heuristic for a suite of examples (quasiperiodic, Lorenz,
and driven damped pendulum dynamics).  We compare our heuristic to
the method of Fraser \& Swinney\cite{fraser-swinney} that chooses the
delay giving the first minimum on a plot of the average mutual
information (AMI) versus $\tau$; see \S\ref{sec:DCE}.
Since our aim is to obtain a dynamical reconstruction that is
diffeomorphic to the full state-space dynamics of the system, we use a
dynamical invariant---the correlation dimension, $d_2$---to compare
the full and reconstructed dynamics.  As described in more detail in
\S\ref{sec:Methods}, this involves computing $d_2$ for embeddings
constructed using the $\tau$ values suggested by the curvature- and
AMI-based heuristics and comparing these to the correlation dimension
of the full state-space trajectory.
The curvature-based heuristic matches the performance of AMI in many
of our examples and outperforms it in others: {\sl e.g.}, when the data
suffers from defects such as noise, limited trajectory length, or
smoothing.  Moreover, the first minima of $\overline{M}(\tau)$ are
generally deeper than those for AMI, making the choice of an
effective $\tau$ more clear.

\section{Parameter Selection in Delay Reconstruction} \label{sec:DCE}

As mentioned above, success in delay reconstruction involves selection
of values for two free parameters: the time delay, $\tau$, and the
embedding dimension, $m$.  A large number of useful heuristics have
been proposed in the literature for estimating these, both separately
and
together.\cite{Gibson92,fraser-swinney,Olbrich97,kantz97,Buzug92Comp,josh-tdAIS,liebert-wavering,Buzugfilldeform,Liebert89,rosenstein94,Cao97Embed,Kugi96,joshua-pnp,KBA92,Hegger:1999yq,kantz97,Holger-and-Liz}
In this paper, we focus on selecting $\tau$, even though the two
parameters have interacting effects.  (Indeed, these interactions
support an elegant method for selecting them at the same time.
\cite{pecoraUnified})

Given a scalar time series $\{x_t, t \in \bN\}$, an $m$-dimensional,
time-delay reconstruction corresponds to the sequence of delay vectors
\begin{equation}\label{eq:delayVector}
	\vec{x}_j=[x_j,~x_{j-\tau}, ~ \dots, ~ x_{j-(m-1)\tau}]^T \in \bR^m
\end{equation}
for a given delay $\tau$.  The time delay or ``lag" $\tau$ defines the
number of steps separating each coordinate.

The theoretical constraints on the time delay are far from stringent,
requiring only $\tau>0$.\cite{sauer91,takens}  This only applies in
the case of infinitely long, noise-free time series and infinite precision arithmetic, however---idealizations that are never realized in
practice.  As a result, the selection of $\tau$ plays an important
role in the practical application of this methodology.
\cite{fraser-swinney,kantz97,Buzug92Comp,liebert-wavering,Buzugfilldeform,Liebert89,rosenstein94,josh-tdAIS}

The fact that the time delay does not play into the underlying
mathematical framework is a double-edged sword.  Because the
theoretical constraints are so loose, there is no practical way to
derive an ``optimal" lag, or even know what criteria such a lag would
satisfy.\cite{kantz97} Casdagli {\sl et al.} \cite{Casdagli:1991a}
provide a discussion of this theory and the impacts of $\tau$ on
reconstructing an attractor for a noisy observation function.
Unfortunately, their discussion gives no practical methods for
estimating $\tau$, even though it does nicely delineate a range of
$\tau$ between {\sl redundancy} and {\sl irrelevance}.  For very small
$\tau$, $x_j$ and $x_{j-\tau}$ are effectively indistinguishable.
This is especially a problem in the presence of noise and finite precision.  In this
situation, the reconstruction coordinates are highly {\sl redundant}:
{\sl i.e.}, they contain nearly the same information about the
system.\cite{Casdagli:1991a,Gibson92}  The implication is that a \emph{very}
small $\tau$ is not a good choice because the additional coordinates
in \eqref{eq:delayVector} add almost nothing new to the model.
Choosing an arbitrarily {\sl large} $\tau$ is undesirable as well,
because it makes the coordinates of the reconstruction become causally
unrelated.  In such a case, the measurement of $x_{j-\tau}$ is {\sl
  irrelevant} in predicting $x_j$.\cite{Casdagli:1991a} Useful
$\tau$ values lie somewhere between these two extrema.

In practice, finding $\tau$ values in this middle ground can be quite
challenging.  The most commonly used method for this involves
computing the time-delayed average mutual information or AMI,
$I(x_j,x_{j-\tau})$, for a range of $\tau$.  Fraser \& Swinney argue
that selecting $\tau$ to give the first minimum of AMI will minimize
the redundancy of the embedding coordinates, thereby maximizing the
information content of the overall delay vector.\cite{fraser-swinney}
This standard method is not without problems.  For some time
series---{\sl e.g.}, processes with memory longer than
$\tau$---$I(x_j,x_{j-\tau})$ does not have a minimum.  This occurs in
any autoregressive process, for instance, and in real-world data as
well.\cite{mytkowicz09,josh-IDA13,jones2018southern} Even if a minimum
exists, it can be shallow, requiring a subjective choice on the part
of the practitioner as to its location, or even its existence.  Noise
and other data issues, such as coarse temporal resolution of the time
series, can also affect the performance of any $\tau$-selection
method, including AMI.

\section{Methods}\label{sec:Methods}

In the absence of formal guidelines for selecting optimal parameters,
standard practice in the delay-reconstruction literature is to
illustrate the usefulness of a parameter-selection method in some
specific context: often, its accuracy in estimating dynamical
invariants,\cite{Buzugfilldeform,rosenstein94,liebert-wavering,Olbrich97}
in or maximizing forecast accuracy from a
reconstruction.\cite{josh-tdAIS,joshua-pnp}

In this paper, we show that our time-delay selection criterion allows
for an accurate calculation of the correlation dimension $d_2$ of an
attractor.
We perform the calculations using the Grassberger-Procaccia
algorithm,\cite{GrassbergerPhysicaD} which approximates $d_{2}$ by
looking for the power law
 \begin{equation}\label{eq:powerlaw}
 	C(\epsilon) \sim \epsilon^{d_{2}}, 
\end{equation}
as the scale parameter $\epsilon \to 0$. Here $C(\epsilon)$ is the
correlation sum
\begin{equation}\label{eq:corrsum}
	C(\epsilon) = \frac{1}{N(N-T)}\sum_{i=1}^N\sum_{j=1}^{i-T} \Theta [\epsilon- ||\vec{x_i} - \vec{x_j}||],
\end{equation}
where $N$ is the number of points in the trajectory,
$\Theta(x)$ is the Heaviside step function, and $T$ is the ``Theiler
window," chosen to ensure that the temporal spacing is large enough to
represent an independently identically distributed sample.  If
\eqref{eq:powerlaw} holds over some sufficiently large
scaling region on a log-log plot, then its slope estimates the
correlation dimension.  A fast algorithm to compute the correlation
sum is available in the TISEAN package.\cite{kantz97,tisean-website}

There are several practical challenges that can affect the computation
of $d_{2}$.  These include selection of the parameters $N$ and $T$, as
well as the range for $\epsilon$.  A persistence-based approach---{\sl
  i.e.}, finding a large range for each parameter that gives
consistent values of the correlation dimension---is perhaps the best
way to make these choices.\cite{kantz97,Holger-and-Liz}  To determine
an appropriate scaling range for \eqref{eq:powerlaw}, the
standard practice is to require that the nearly linear relationship
between $\log (C(\epsilon))$ and $\log(\epsilon)$ exist over a {\sl
  considerable} range of scales.  This is necessarily subjective and
makes the process challenging, if not impossible, to automate.  We
discuss the specifics of our approach to this problem in the Appendix.

\section{Results}\label{sec:Results}

In this section, we demonstrate the performance of our curvature-based
heuristic using three example systems: quasiperiodic motion on a
two-torus, the classic Lorenz system, and a driven damped pendulum.
In each case, we compute a representative trajectory and choose one of
the state-space variables as the measurement function for the
time-delay reconstruction.  We then compute the mean Menger curvature
$\overline{M}(\tau)$ of a 2D delay reconstruction of those data for a
range of values of $\tau$, using a {\tt C++} implementation of
\eqref{eq:menger}.  We choose the time-delay, $\tC$, as the first
minimum of that curve.  For the purposes of comparison, we also
compute the average mutual information profile $I(x_{t}; x_{t-\tau})$
using the {\tt mutual} command in the {\tt TISEAN}
package,\cite{tisean-website,kantz97} choosing $\tAMI$ at the first
minimum of that curve.  For the Lorenz and pendulum examples, we use
{\tt TISEAN}'s {\tt d2} to calculate the correlation dimension of the
full state-space trajectory and compare it to the correlation
dimension of embeddings constructed with $\tC$ and $\tAMI$.  Details
of these computations are given in the Appendix, including the
correlation sum plots and a discussion of the nuances of choosing the
scaling region.
For the two-torus, we compare the calculated $d_2$ value of the
embedding to the known dimension of the system.  We also explore and
discuss the effects of noise, low-pass filtering, and data length
using the Lorenz and pendulum systems.

\subsection{Quasiperiodic Dynamics on a Two Torus} \label{sec:Torus}

We begin our discussion with a simple quasiperiodic
system---incommensurate rotation on a two-torus with the flow:
\begin{align}\label{eq:torus}
x &= (R + r\cos(2\pi\ t))\sin(2\pi\phi t), \nonumber \\
y &= (R + r\cos(2\pi\ t))\cos(2\pi\phi t),\\
z &= r\sin(2\pi\ t) \nonumber .
\end{align}
Here, we use the radii $R = 1$ and $r = 0.2$ and frequency ratio $\phi
= \tfrac12(1+\sqrt{5})$, the golden ratio.  For this example, we
generate a time series of 200,000 points with a time spacing of
$0.001$.  The experimental time series, $x(t)$, is the first variable of~(\ref{eq:torus}) as a function of time.
 
\begin{figure}[htbp]
	{
		\includegraphics[width=0.8\columnwidth]{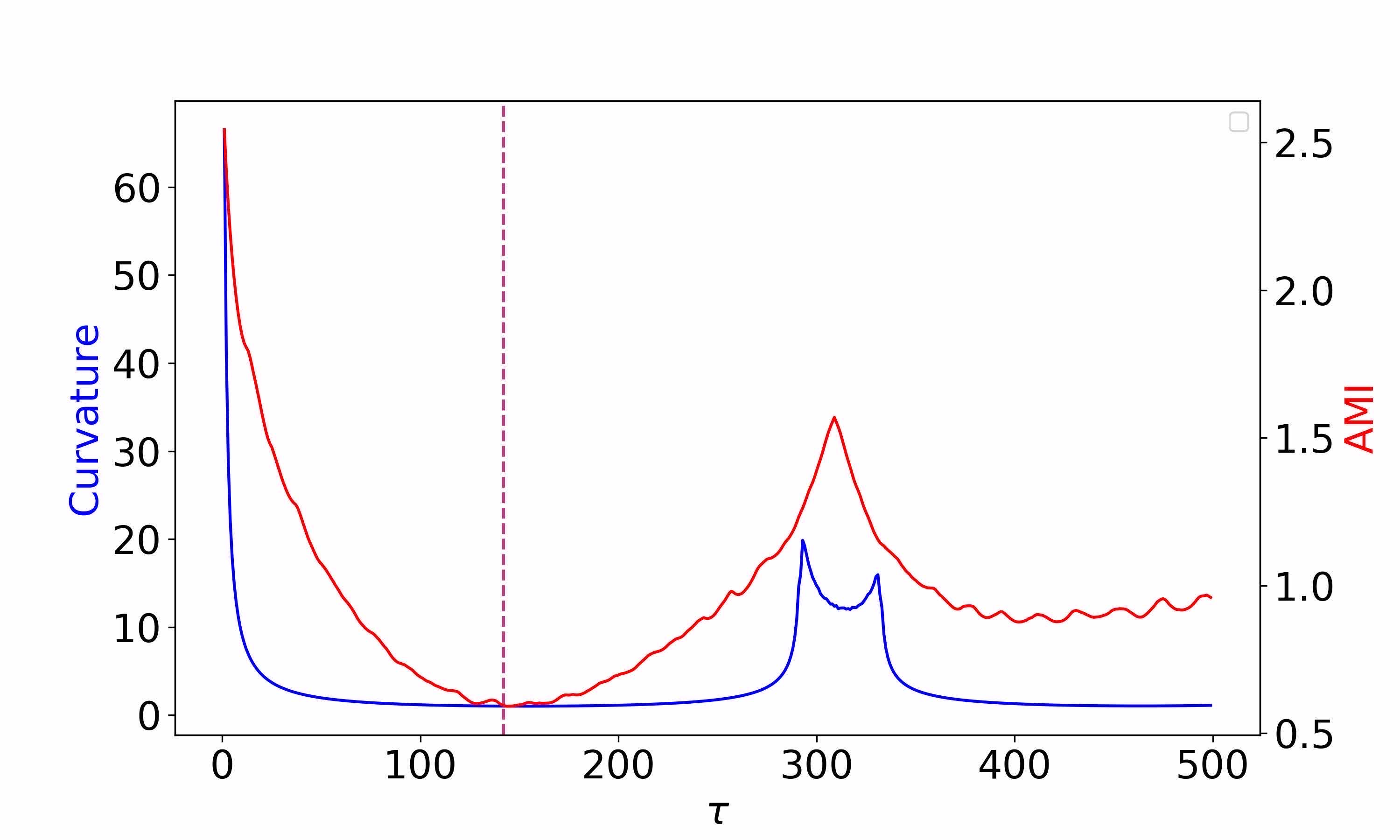}
	} \\
	{
		\includegraphics[width=0.8\columnwidth]{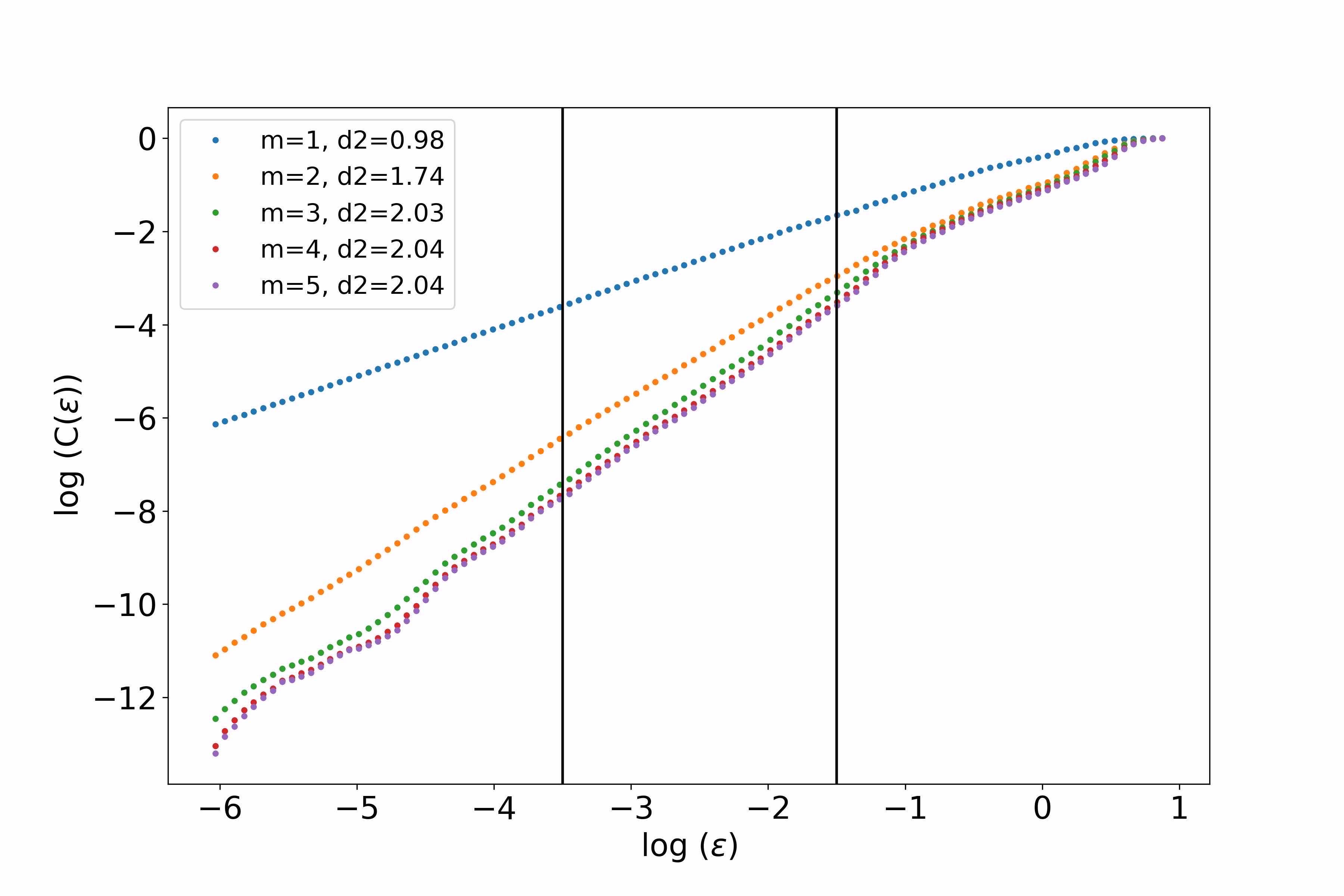}
	}
	\caption{Top: Mean curvature (blue) and average mutual
          information (red) profiles as a
          function of time delay, for 2D reconstructions of the
          two-torus dynamics from $x(t)$. 
          Bottom: Correlation sums \eqref{eq:corrsum} for $m$-dimensional
          delay reconstructions with $\tau=142$. The scaling region between the two vertical
          lines is used to compute
          $d_2$.}  
      \label{fig:torus_measures_and_d2}
\end{figure}

The mean curvature and AMI plots for 2D reconstructions of these data
are shown in Fig.~\ref{fig:torus_measures_and_d2}(a).
The first minima of the two profiles fall at the same value: $\tAMI =
\tC = 142$.  Fig.~\ref{fig:torus_measures_and_d2}(b) shows the
correlation sum plots for reconstructions constructed with this $\tau$
value and $m$ ranging up to five, the value generically sufficient for
embedding a two-dimensional invariant set, according to
Takens.\cite{takens} The correlation dimension converges to 2.04 for
$m \ge 4$, close to the true dimension of the two-torus.
Interestingly, the profile of the mean curvature has a wide flat
region around its first minimum, suggesting that any value $\tC \in
[100, 200]$ could be appropriate.  Indeed, the correlation dimension
is unchanged for time delays across this interval.  It is often
advantageous, for a variety of
reasons,\cite{joshua-pnp,Casdagli:1991a,josh-tdAIS,Olbrich97} to
choose the time delay as small as possible to maintain the accuracy of
the reconstructed dynamics.  The fact that the curvature suggests a
wide range for $\tau$ would allow a practitioner to exploit this
advantage by selecting a smaller $\tau$; by contrast, the AMI curve
would constrain that choice to a single, larger value.

\subsection{Classic Lorenz Attractor}\label{sec:Lorenz}

\begin{figure}
	\includegraphics[width=0.8 \linewidth]{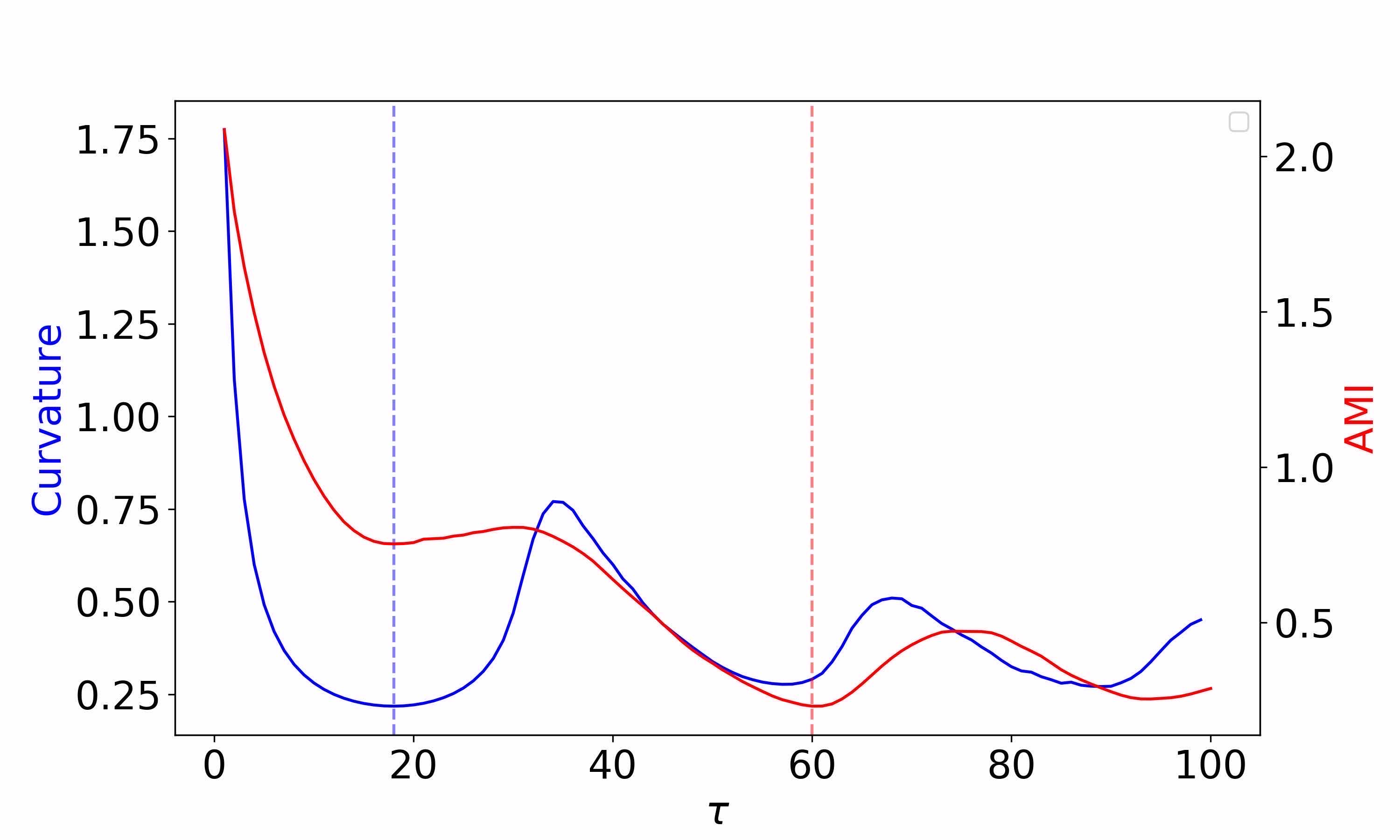}
\caption{Mean curvature and AMI profiles for the Lorenz data: 
the $x$ coordinate of Fig.~\ref{fig:lorenz}(a).}
\label{fig:lorenz_heuristic_curves}
\end{figure}

As a second example, we consider the canonical Lorenz system,\cite{lorenz}
\begin{align}
\dot{x}&= \sigma(y - x), \nonumber\\
\dot{y} &= x(\rho - z) - y,\\
\dot{z} &= xy - \beta z \nonumber,
\end{align}
for the standard parameters $\sigma = 10$, $\beta = \tfrac83$,
and $\rho = 28$.  We use a fourth-order Runge-Kutta method with a fixed
time step of 0.01 and integrate the system to $t=1000$.  Discarding
the initial transient $t < 100$, we obtain a time series of length
90,000.  As before, we take $x(t)$ to be the measurement function.
Fig.~\ref{fig:lorenz_heuristic_curves} shows the mean curvature and
average mutual information plots for these data.
The first minima of both curves fall near $\tau = 18$.  Note, though,
that the minimum in the curvature profile appears significantly
deeper.  Indeed, based on this AMI profile, it is quite likely that a
practitioner would actually choose $\tAMI = 60$.  

To formalize the notion of depth of a minimum, we define
\begin{equation}\label{eq:depth}
	\Delta = (H_{max} - H_{min})/H_{max},
\end{equation}
where $H_{min}$ is
the height of the curve at that minimum and $H_{max}$ the height at
the subsequent (here, higher $\tau$) maximum.  For the first minimum
in the AMI profile, $\Delta=0.074$; for $\overline{M}$, $\Delta=0.72$.  The
deeper minimum of the $\overline{M}$ curve is a significant advantage in
this regard, as it makes the choice far more clear. Nevertheless, both
values of $\tau$ yield correlation dimensions that are good matches
to the correlation dimension of the full 3D trajectory ($d_2 \approx 2.05$): 
for $\tC$, we find $d_2 = 2.04$, and for $\tAMI$, $d_2 = 2.07$.\footnote
{More details are in the Appendix and Fig.~\ref{fig:lorenz_d2}.}

Even though, in this particular case, the two embeddings give
nearly the same dimensions, a comparison of panels (b) and (d) of
Fig.~\ref{fig:lorenz_curvature_heatmaps} suggests other issues that
are potential problems. In particular, the smaller $\tau$ produces a
far less geometrically complicated reconstruction.  The folds and
kinks in the $\tau=60$
embedding---Fig.~\ref{fig:lorenz_curvature_heatmaps}(d)---could be
problematic if noise were present. This is precisely where we turn our
attention next.

\subsection{Noisy Lorenz Dynamics}\label{sec:noisyLorenz}

To explore the effects of noise on the choice of time delay, we use
the same Lorenz trajectory as in \S\ref{sec:Lorenz} but add
\textit{iid} noise to each point in the trajectory time series with a
uniform distribution of amplitude $0.1 x_{max} = 1.922$, {\sl i.e.}, a
noise-to-signal ratio of $10\%$. For a fair comparison of
the reconstruction to the full, 3D trajectory, it is important to add
noise to each component.  The correlation
dimension calculated from the full, noisy trajectory is $d_2 \approx
2.30$,\footnote
{The scaling
regions for the correlation dimension calculations need to be chosen
carefully in such situations to factor in the noise levels; see the
Appendix for more discussion.}
slightly larger than for the noise-free case, as one would expect.%

\begin{figure}
	\includegraphics[width=0.8 \linewidth]{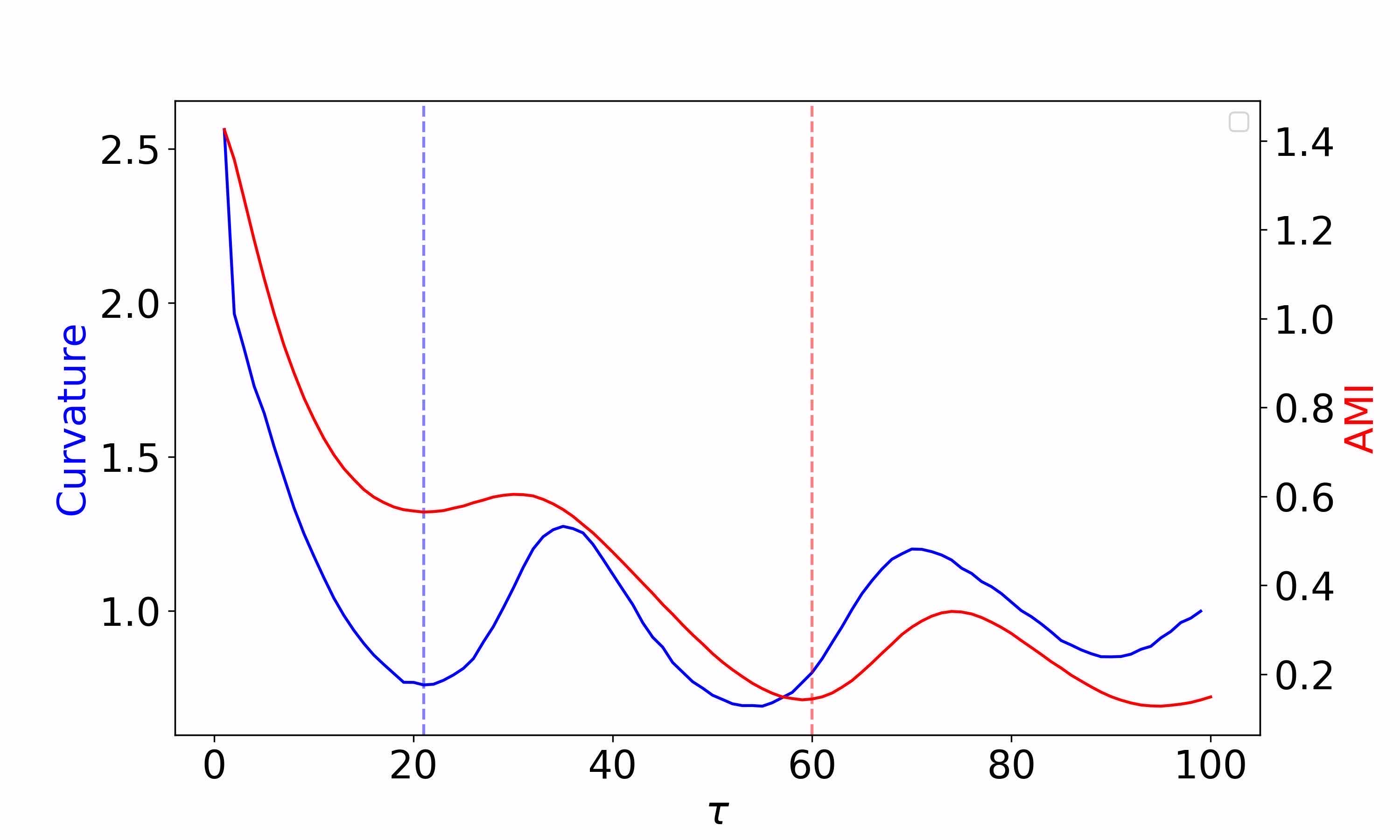}
\caption{Mean curvature and AMI profiles for the noisy Lorenz data. Compare to 
the curves without noise in Fig.~\ref{fig:lorenz_curvature_heatmaps}.}
\label{fig:lorenz_noise_heuristics}
\end{figure}

The AMI and curvature profiles for 2D reconstructions of these noisy
data are shown in Fig.~\ref{fig:lorenz_noise_heuristics}.  
As in the noise-free case, the AMI profile has a comparatively shallow
first minimum ($\Delta=0.066$), now at $\tau = 20$, and a more
well-defined minimum ($\Delta=0.59$) at $\tau = 60$.
This suggests the choice $\tAMI = 60$.  Note that the first minimum
of $\overline{M}$, at $\tC= 21$, is still well-defined ($\Delta=0.41$).
The corresponding correlation dimensions\footnote
{See Fig.~\ref{fig:lorenz_noise_d2} in the Appendix.}
are $d_2 = 2.28$ for $\tC$ and $3.83$ for $\tAMI$---the latter far
higher than the correct value.  These results suggest that the
curvature-based heuristic works well in the face of noise, perhaps
because the smaller $\tau$ that it finds produces less overfolding of
the reconstructed trajectory.
To explore this further, we varied the
noise-to-signal ratio from $0.001$ to $0.1$, and found that $\tC$
remains relatively steady in the range $[18,21]$, producing a
dimension consistent with the full trajectory. By contrast, the value
of $d_2$ for the AMI reconstruction steadily diverges from the correct
value as the noise grows.  This adds to our confidence about the
robustness of the curvature-based heuristic with respect to noise.

\subsection{Driven Damped  Pendulum}\label{sec:Pendulum}

\begin{figure*}
		\subfloat[]{
		\includegraphics[width=0.32\linewidth]{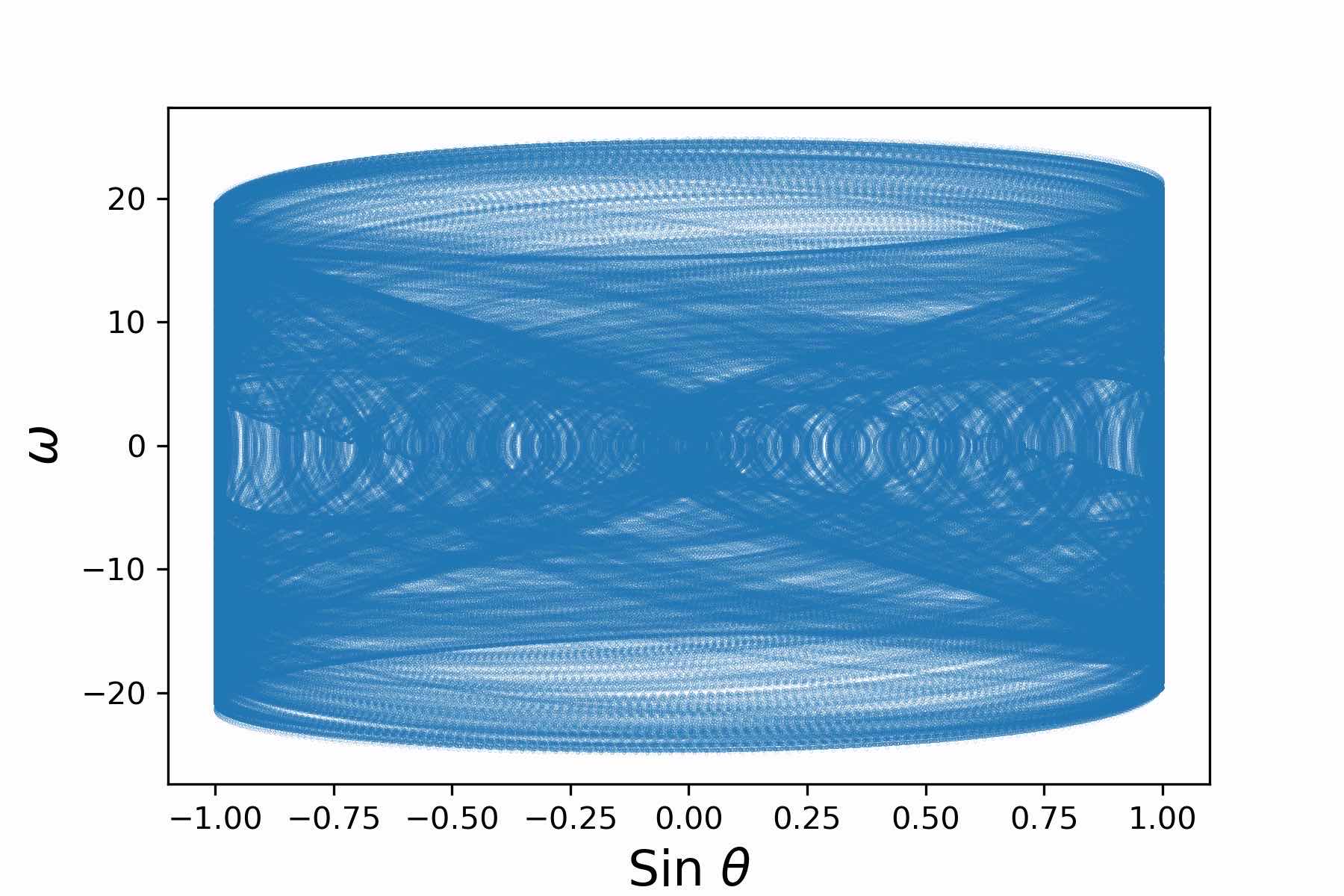} 
		}
		\subfloat[]{
		\includegraphics[width=0.32\linewidth]{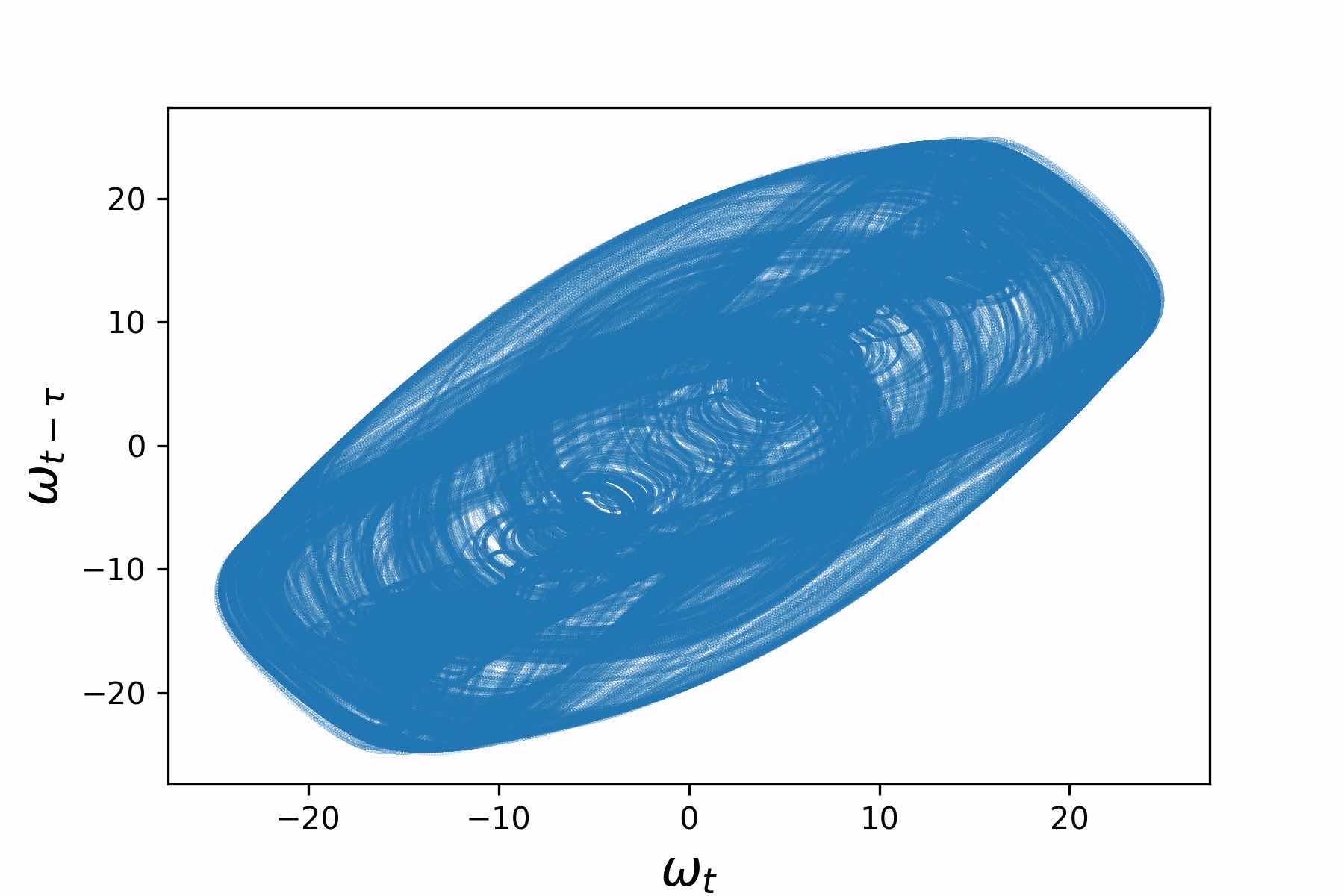}
		}
		\subfloat[]{
		\includegraphics[width=0.32\linewidth]{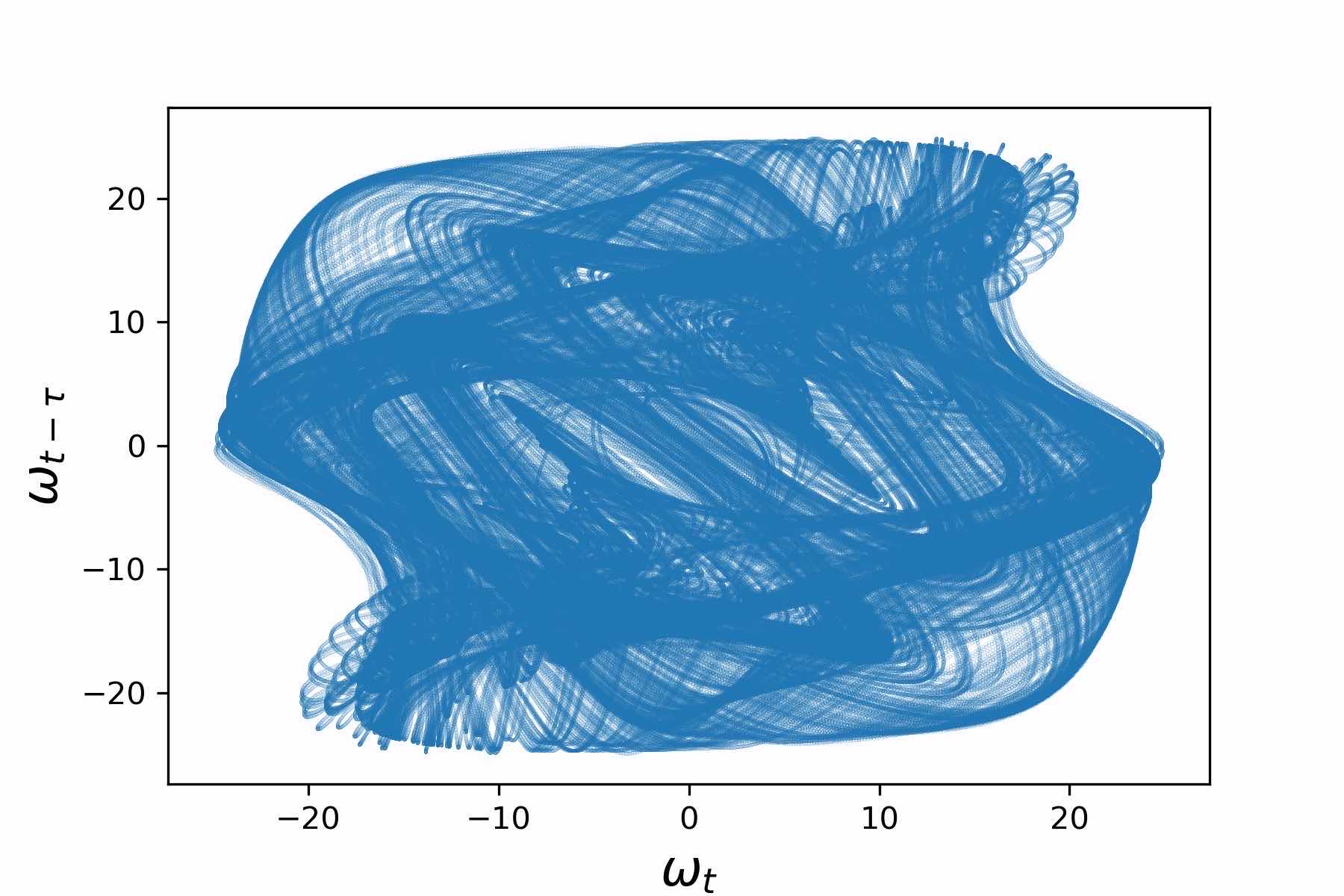}
		}
	\caption{The driven damped pendulum: (a) a projection of the
	full state space onto $(\sin \theta,\omega)$; delay
	reconstructions of $\omega$ for (b) $\tau = 120$; (c)
	$\tau=250$. }
	\label{fig:pendulum_recons}
\end{figure*}

\begin{figure}
	\includegraphics[width=0.8 \linewidth]{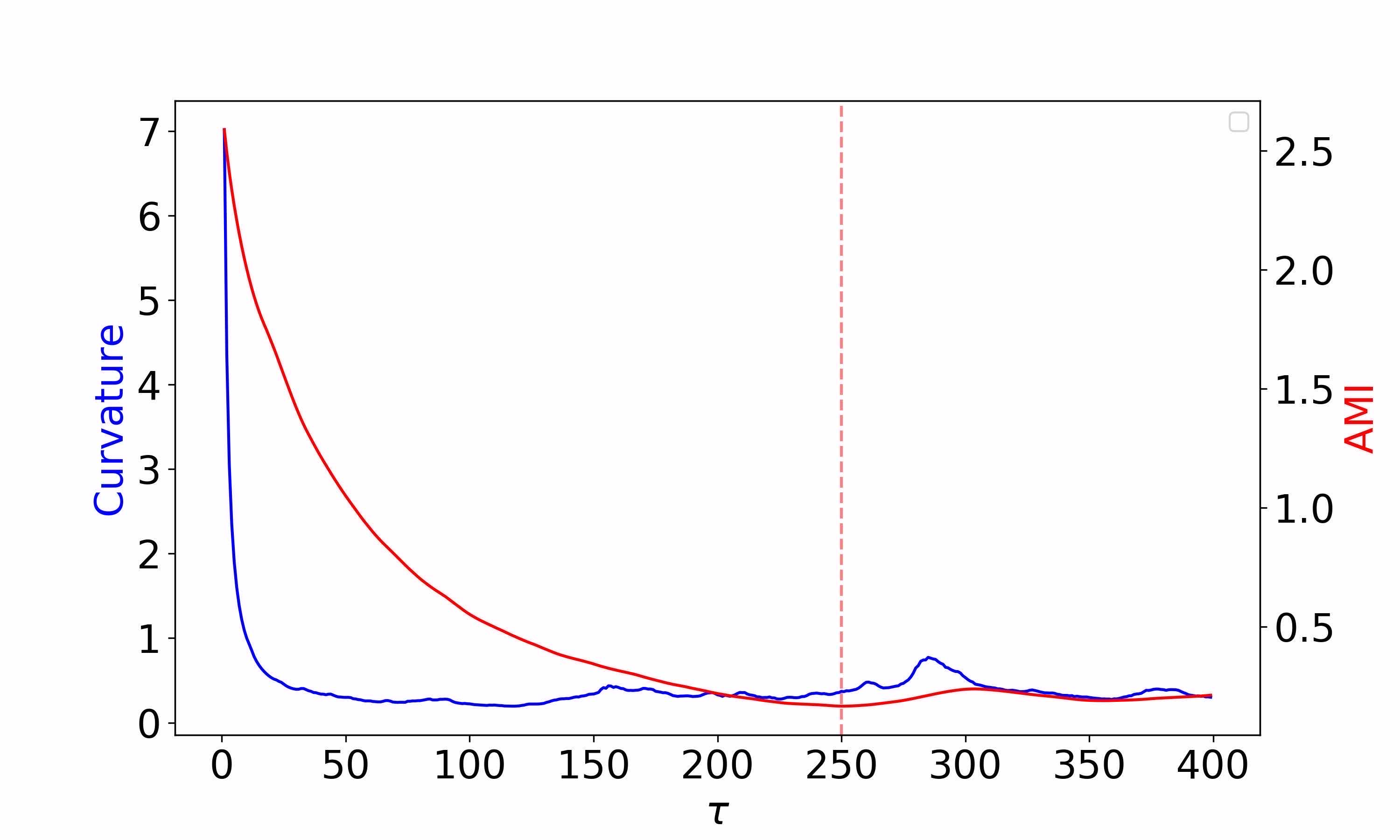}
\caption{Mean curvature and AMI profiles for the driven damped
  pendulum using $\omega$ as the measurement function.}
\label{fig:pendulum_heuristics}
\end{figure}

In this section, we consider a pendulum with natural frequency $\nu_0$
that is subject to linear damping and a time-periodic force:
\begin{equation}\begin{split}\label{eq:Pendulum}
	\dot{\theta} &= \omega, \\ \dot{\omega} &= -\beta\omega
	- \nu_0^2 \sin(\theta) +A \cos(\alpha\ t),\\
\end{split}\end{equation}
The coordinates of the 3D, extended phase space $\mathbb{S} \times
\mathbb{R} \times \mathbb{S}$ are angle $\theta$, angular momentum
$\omega$, and time $t$.  To fix parameters, we choose $\nu_0^2 = 98$,
damping $\beta = 2.5$, and a driving force with amplitude $A = 91$ and
frequency $\alpha = 0.75 \nu_0$.  This system has a chaotic attractor.
As in the Lorenz example, we solve the system \eqref{eq:Pendulum}
using fourth-order Runge-Kutta, now with time step of $0.001$.
Discarding the first $10^5$ points to eliminate transient behavior, we
keep a time series of $10^6$ points.  To avoid issues with periodicity
in $\theta$ and $t$, we project the time series onto the three
variables $\{\sin(\theta(t)), \omega(t), \sin(\alpha t)\}$, as seen in
Fig.~\ref{fig:pendulum_recons}(a).
For the time-delay reconstruction experiments, we take $\omega$ as the
measurement function.  Profiles of the mean curvature and AMI of this
signal are shown in Fig.~\ref{fig:pendulum_heuristics}.
The former has a broad plateau in the range $50 \lesssim \tau \lesssim
120$, which again provides the flexibility to choose the lowest
possible $\tau$ that successfully reconstructs the dynamics.  The AMI
has a first minimum at $\tAMI = 250$.  For this $\tau$ value, as well
as for values across the range $50 \lesssim \tau \lesssim 120$, the
calculated correlation dimension $d_2=2.22$, which matches the
correlation dimension of the full state-space dynamics.  In this case,
the curvature heuristic appears to match the performance of AMI,
perhaps because $d_2$ is relatively insensitive to the choice of
$\tau$, as in the Lorenz example.  However, there are significant
geometrical differences between the resulting 2D reconstructions; see
Fig.~\ref{fig:pendulum_recons}(b)-(c).  The larger $\tAMI$ produces an
overfolded reconstruction, increasing the curvature along the
trajectory.  This can, as demonstrated in the previous example,
increase noise sensitivity.

\subsection{Other Data Effects}\label{sec:data-effects}

Data limits---shorter traces or coarser temporal sampling---are
another practical issue in delay reconstruction.  To explore the
effects of data length on the curvature-based heuristic, we repeat
the pendulum experiment with a shorter trajectory, keeping only the
first $200,000$ points, one fifth of the previous time series. 
The resulting AMI and mean-curvature profiles (not shown) are essentially identical
to those in Fig.~\ref{fig:pendulum_heuristics}, and
the correlation dimension for the $\tC$ reconstruction ($d_2 =
2.20$) is close to that of the full dynamics ($d_2 = 2.18$);
however, for a reconstruction using $\tAMI$, $d_2$ increases to $2.35$.\footnote
{See Figs.~\ref{fig:pendulum_d2} and \ref{fig:pendulum_200K_d2} in the
Appendix for the calculations.}
This suggests that the effects of the larger $\tau$
can be more significant for smaller data sets.  Indeed, if there are
fewer points, those that are artificially close, due to overfolding,
will have a larger effect on the correlation function.

\begin{figure}
	\includegraphics[width=0.8 \linewidth]{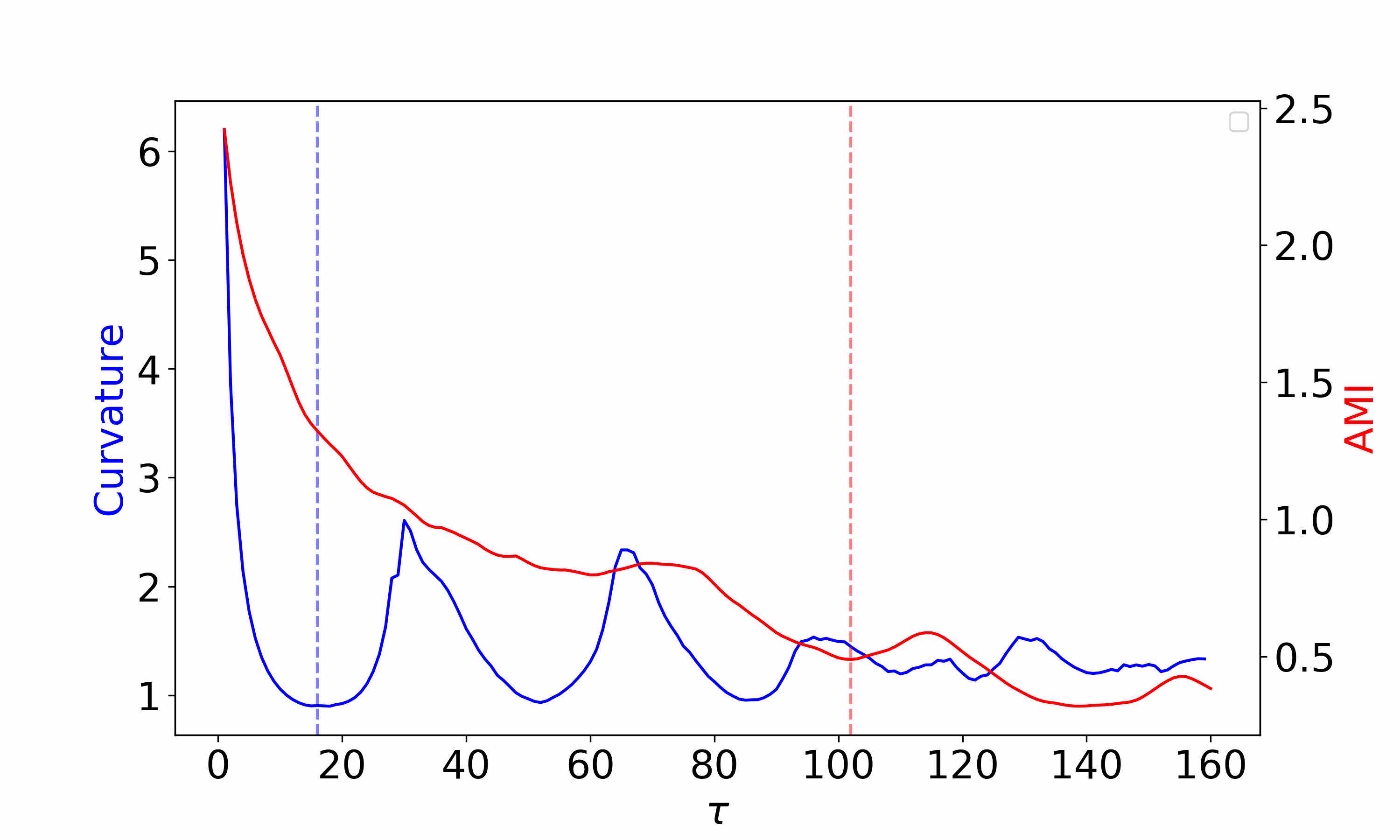} 
\caption{Mean curvature and AMI profiles for the filtered Lorenz data.
Compare to Fig.~\ref{fig:lorenz_heuristic_curves}.}
	\label{fig:lorenz_filtered_heuristics}
\end{figure}

Another issue that arises in the practice of nonlinear time-series
analysis is data smoothing.  This often occurs during the measurement
process\cite{jones2018southern} or in data-processing
pipelines,\cite{Pennekamp2019,mytkowicz09,josh-IDA11} but can also
occur naturally through, for example, diffusive
processes.\cite{jones2017water}  
To explore the potential effects of
this upon the different $\tau$-selection heuristics, we ran the Lorenz
time series (\S\ref{sec:Lorenz}) through a moving average filter on a
data window $[j-60,j+59]$ for the $jth$ point:
\[
	\tilde{x}_{j} = \frac{1}{120}\sum_{i=-60}^{i=59} x_{j+i} .
\]
The mean curvature and AMI profiles for the smoothed time series
$\tilde{x}$ are shown in Fig.~\ref{fig:lorenz_filtered_heuristics}.
The first clear minimum for the curvature moves slightly, to $\tC =
16$, with a depth $\Delta = 0.65$.  The AMI profile, though, is
significantly distorted from that seen in the non-filtered data of
Fig.~\ref{fig:lorenz_heuristic_curves}: there is now no minimum at
$\tau = 18$ and only a weak one at $\tau = 60$ ($\Delta = 0.05$).
This leaves the minimum at $\tau = 102$ ($\Delta = 0.17$) as the
reasonable choice for $\tAMI$.

This deformation makes sense: a moving-average filter decreases the
independence of neighboring points, thereby increasing the AMI and
obscuring any minima that occur at lower $\tau$ values.  As in the
previous case of limited data, the larger $\tAMI$ creates problems
with the dimension calculation: $d_2$ for the $\tAMI$ embedding is
$1.89$---a significant under-estimate.  In contrast, $d_2 = 2.11$ for
the $\tC$ embedding, which is still close to the value $d_2 = 2.05$ of
the full, filtered dynamics.\footnote
{Details presented in Fig.~\ref{fig:lorenz_filtered_d2} of the
Appendix.}
This is an encouraging result; obtaining a reconstruction that matches
the underlying dynamics, even though the data are smoothed, is quite
useful.

Over- and under-sampling---{\sl i.e.}, temporal spacing between data
points that is far smaller or far larger than the time scales of the
dynamics---are also issues in the practice of nonlinear time-series
analysis.  Under-sampled data are, of course, a challenge to any
method and there is very little recourse in that situation, as one
should not make up data.  Over-sampling causes different problems.
For the curvature heuristic, points spaced too closely along the
trajectory will be nearly co-linear, which can cause numerical issues
in ~\eqref{eq:menger}.  AMI, in contrast, is relatively immune to this
problem; oversampling simply moves its first minimum to a higher
$\tau$ value.  One can determine whether over-sampling is an issue
using standard best practices: {\sl viz.,} repeating the analysis with every
$n^{th}$ point and observing whether the results
change.\cite{kantz97,Holger-and-Liz}

\section{Conclusions and Future Work}\label{sec:Conclusion}

The curvature heuristic that we have proposed is the following: choose
a time-delay $\tau$ to be the value giving the first local minimum of
the average of the local Menger curvature, \eqref{eq:localC}, along a
segment of a 2D delay reconstruction of a scalar time series
$\{x_j\}$.  The set of experiments presented in the previous section,
which are summarized in Table~\ref{tab:d2_results}, suggest that this
heuristic is useful for selecting an appropriate embedding delay.  In
particular, in every case, the $\tau$ value suggested by that
heuristic gives a correlation dimension from a time-delay embedding
({\sl i.e.}, a reconstruction with a sufficiently large dimension $m$)
that agrees with the dimension of the full attractor---within
reasonable error bounds.

\begin{table}[htbp]
	\begin{tabular}{cc|c|c|c}
		& & Full Dynamics & Curvature  &AMI \\
 		\hline
		\multirow{2}{*}{Two-Torus} &	True dynamics & 2.0  & 2.04  & 2.04 \\
									& $\tau$ & - & 142 & 142  \\
		\hline 
		\multirow{2}{*}{Lorenz} &	$d_2$& 2.05  & 2.04  & 2.07 \\
									& $\tau$ & - & 18 & 60  \\
		\hline 
		\multirow{2}{*}{Lorenz + Noise} &	$d_2$& 2.30  & 2.28  & 3.83 \\
									& $\tau$ & - & 21 & 60  \\	
		\hline 
		\multirow{2}{*}{Pendulum (long)} &	$d_2$& 2.22  & 2.22  & 2.21 \\
									& $\tau$ & - & 120 & 250  \\
		\hline 
		\multirow{2}{*}{Pendulum (short)} &	$d_2$& 2.18  & 2.20  & 2.35 \\
									& $\tau$ & - & 120 & 250  \\
		\hline 
		\multirow{2}{*}{Lorenz + Filtering} &	$d_2$& 2.05  & 2.11  & 1.89 \\
									& $\tau$ & - & 16 & 102  \\	
		\hline 
	\end{tabular}
	\caption{The dimensions for various systems and their 2D delay
          reconstructions using the time delay chosen by mean
          curvature and AMI heuristics.}
	\label{tab:d2_results}
\end{table}

Just like the conventional AMI heuristic, which is based on averaging
the mutual information, this minimum can be computed before one knows
the correct embedding dimension.  The curvature of a trajectory in a
2D reconstruction is a geometrical signal that can be used to diagnose
problems with the choice of time delay $\tau$.  For example, when
$\tau$ is too small, the reconstructed trajectory lies near the
diagonal---it is not properly unfolded.  This leads to sharp reversals
in the 2D reconstruction near local maxima of the time series, which
locally increases their curvature.  Similarly, when $\tau$ is too
large, a 2D delay reconstruction tends to overfold: such folds again
result in larger curvature on parts of that reconstruction.

The curvature heuristic appears, from our examples, to provide some
advantages regarding the choice of minimum: sometimes, a more-distinct
first minimum than that exhibited by the AMI; in other cases, a wider
first minimum that allows useful flexibility in the choice of $\tau$.
This is important, as ambiguities in identifying the first minimum of
AMI can lead to overfolded reconstructions.  We have also shown that
the curvature heuristic is robust to noise, while the AMI heuristic
can be less so.  The curvature heuristic also appears to be less
influenced by shorter data lengths and low-pass filtering.  In
practice, limited data, as well as data that has been aggregated in
some way, are quite common; in these cases, our heuristic may prove to
be useful.

In the future, we plan to study in more detail the full distribution
of the Menger curvature along trajectory sequences.  From the point of
view of Frenet, a smooth trajectory defines a curve in phase space
that generically has an associated orthonormal frame and a set of
generalized curvatures.\footnote
{For example, in 3D the second curvature is the torsion.}
It is possible that additional information can be gleaned from the
full statistics of one or more of these curvatures---information that
would lead to a better criterion than the average.  It would also be
useful to evaluate the effectiveness of the curvature in selecting a
time delay for other purposes, such as in nonlinear forecasting.  Akin
to Pecora {\sl et al.},\cite{pecoraUnified} one could perhaps leverage curvature-based
metrics to select $m$ and $\tau$ simultaneously.

\begin{acknowledgments}
The authors acknowledge support from National Science Foundation Grant CMMI-1537460. In addition, JDM was partially supported by NSF grant DMS-1812481. 
JG was supported by an Omidyar and Applied Complexity Fellowship at the Santa Fe Institute. The authors would like to thank Holger Kantz and Erik Bollt for numerous useful discussions.
\end{acknowledgments}
\section*{Data Availability}
The data that supports the findings of this study are available within the article.  

\bibliography{master-refs}

\clearpage
\appendix
\section*{Appendix: Correlation Dimension Calculations}\label{sec:Appendix}


Determination of the correlation dimension from the correlation sum
plots produced by TISEAN's {\tt d2} tool requires care in choosing the
scaling region, a significant straight segment on the log-log plot.
We make an initial choice by hand, then use the python {\tt polyfit}
tool to fit a line to that segment, and iteratively expand or contract
its width so as to minimize the fit error.  In the plots in this
Appendix, the chosen regions are delineated with vertical lines.

For validation purposes, we compare the correlation dimensions of the
true and reconstructed dynamics.  This requires that one choose the
dimension $m$ of the reconstruction, and a too-small choice for that
value will cause the {\tt d2} results to be incorrect.  Faced with
data from an unknown system, standard practice entails using a
heuristic like the false near neighbor method \cite{KBA92} to estimate
$m$, then validating that choice by repeating the $d_2$ calculation
over a range of $m$ values and looking for convergence.  For the
examples in this paper, where we know the equations, that is
unnecessary; we can perform these comparisons with the full
trajectories and correct embeddings: i.e., reconstructions with $m =
2d+1$, where $d$ is the state-space dimension of the full system (this
is generically a sufficient condition for an embedding\cite{takens,
sauer91}).  In the plots in this Appendix, we repeat the {\tt d2}
calculation for reconstructions over the range $1 \leq m \leq 2d+1$,
where $d$ is the (known) state-space dimension of the system, in order
to understand how the {\tt d2} results change, but we only report the
value for $m = 2d+1$ as the correct correlation dimension.

Error estimates for $d_2$ are notoriously problematic, since these
complex algorithms have many free parameters and subjective choices
about interpretation of the plots.\cite{kantz97} In our examples, the
computed values of $d_2$ are relatively insensitive to the choice of
the scaling interval, provided the initial hand-selected choice falls
within the linear region.  Upon varying the interval by 20\%, we
typically find that the changes in dimension are about 0.2\%

The Theiler window, $T$ in \eqref{eq:corrsum}, is another important
free parameter in {\tt d2} computations.  Recall that the correlation
sum estimates the number of neighboring points present in an
$\epsilon$-ball around sample points on the attractor.  By default,
these also include the points located along the immediate trajectory
of the attractor at the sample point: that is, its immediate temporal
neighbors, in forward and backwards time.  The $d_2$ estimate will 
be biased if the points in the
$\epsilon$-ball consist primarily of these neighbors;
indeed, if {\sl all} of the points in the ball are
immediate temporal neighbors, the correlation dimension algorithm will
return $d_2=1$.  The Theiler window addresses this issue by defining a
set of temporal neighbors around each sample point that will be
ignored in the correlation sum.

We choose the Theiler window so as to exclude points along the
immediate trajectory segment with a span given by the maximum of the
scaling region used in the calculation.  This is a circular problem,
since the choice of the Theiler window affects the scaling region and
its limits.  We solve this iteratively, choosing a small Theiler
window as a starting estimate to get a first approximation of the
correlation sum plots with a visible scaling region, setting the
scaling region limits, updating the Theiler window, as stated, to the
upper limit, recomputing the correlation sum plots, and repeating
until the results stabilize.

\begin{figure*}[htb]
		\subfloat[Full Lorenz dynamics]
{
			\includegraphics[width=0.33\linewidth]{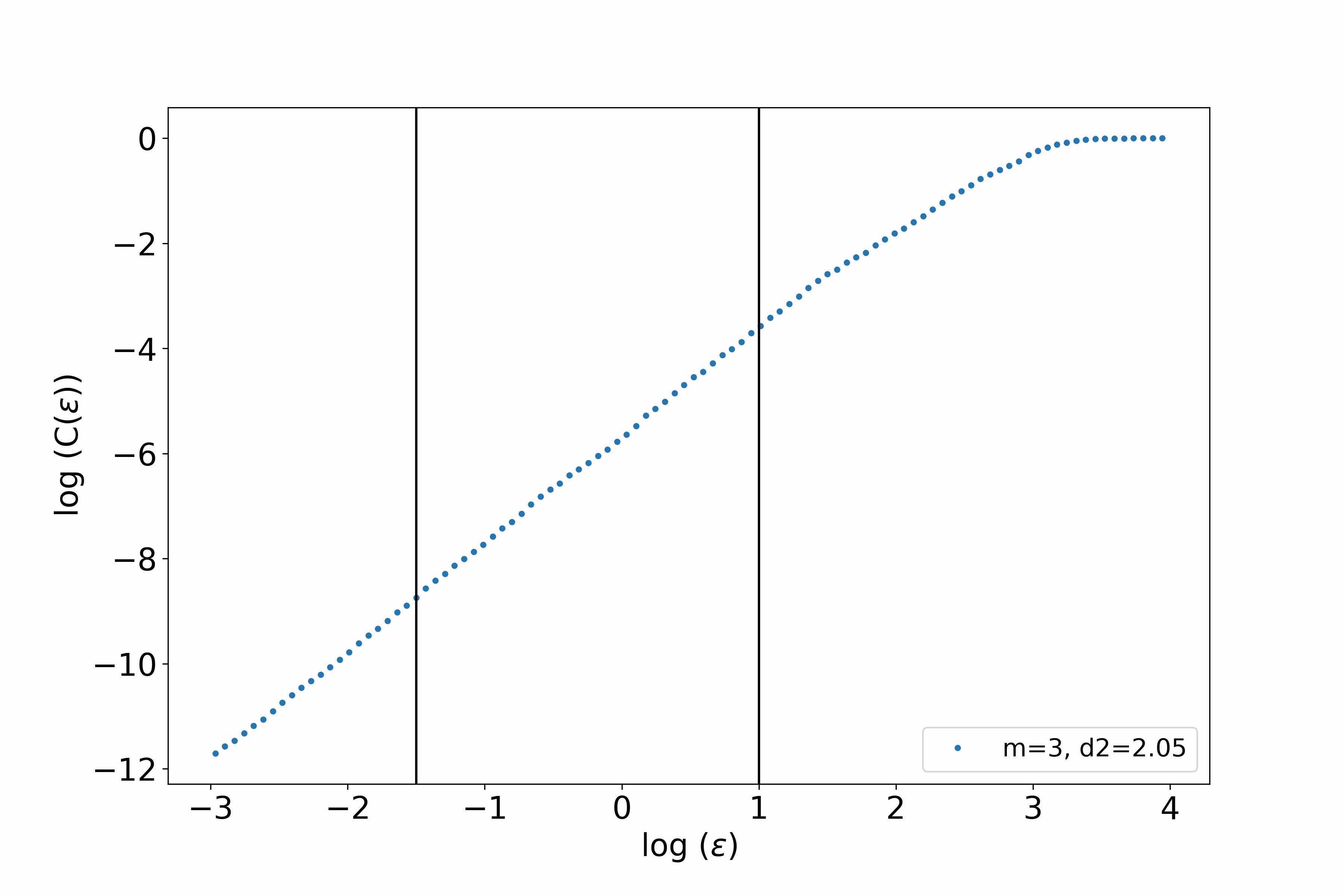}
		}
		\subfloat[$\tau=18$]
{
			\includegraphics[width=0.33\linewidth]{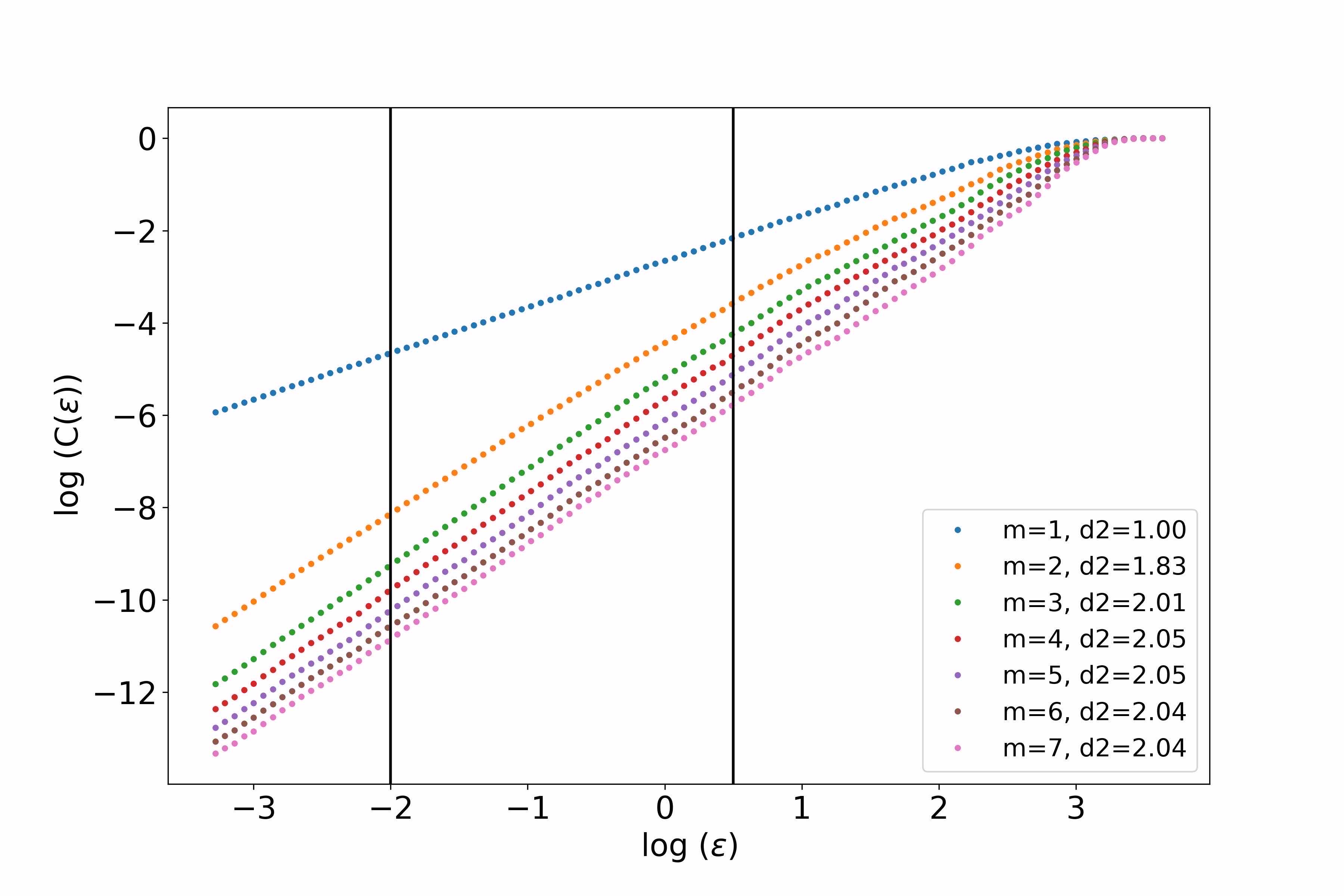}
		}
		\subfloat[$\tau=60$]
{
			\includegraphics[width=0.33\linewidth]{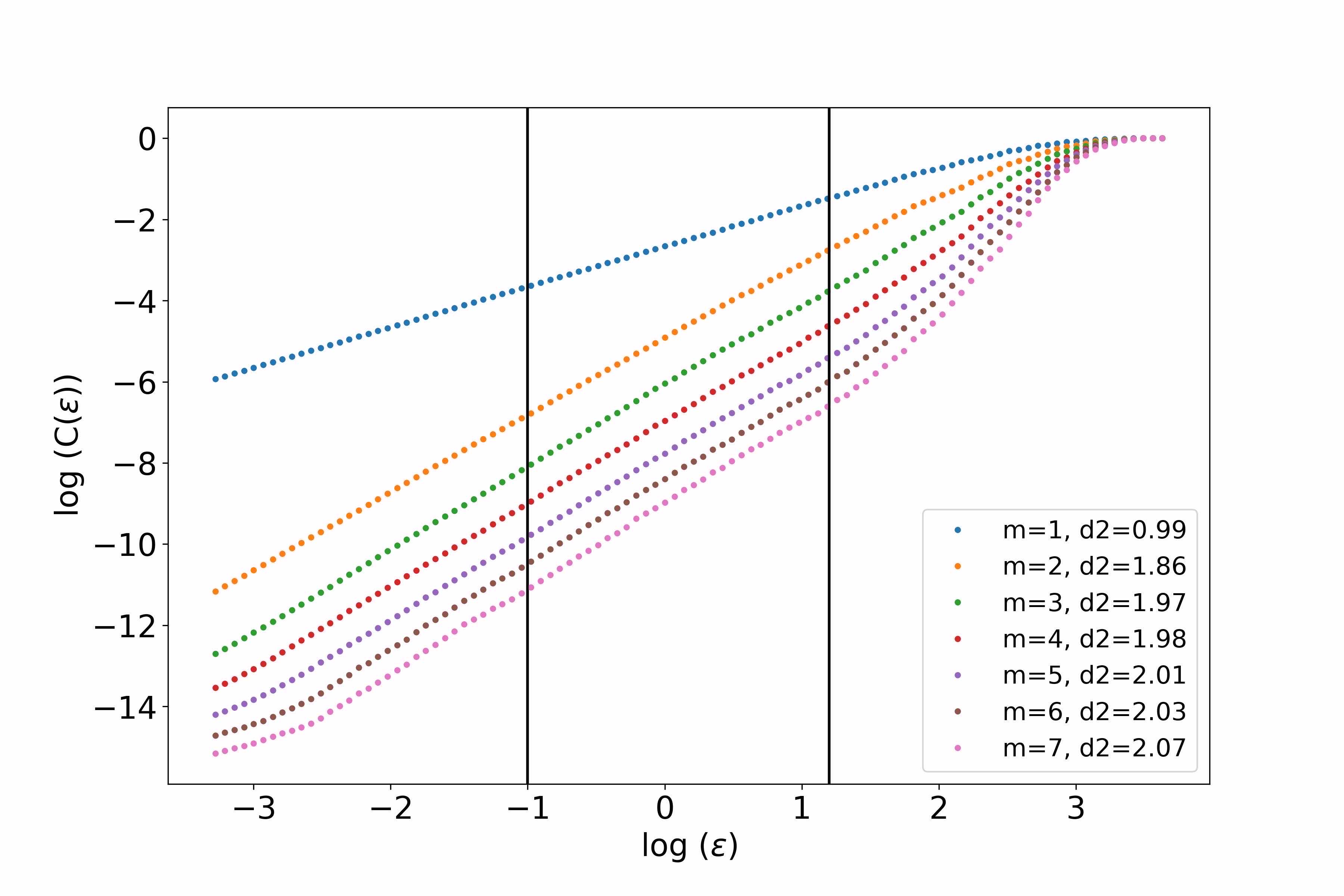}
		}
	\caption{Log-log plots of $C(\epsilon)$ and computed
          correlation dimension of the Lorenz data set for the full
          dynamics, giving $d_2 = 2.05$, and using $\tau=18$---the
          value suggested by the mean curvature heuristic---and
          $\tau=60$, the first clear minimum of AMI for that data
          set.}  \label{fig:lorenz_d2}
\end{figure*}

The correlation plots for the Lorenz data of \S\ref{sec:Lorenz} are shown
in Fig.~\ref{fig:lorenz_d2}. The dimension calculation yields
an estimate of $d_2 = 2.05$ over the scaling region $\log(\epsilon)
\in [-1.5, 1]$ with a Theiler window of 271 data points.
It is accepted that the Hausdorff dimension of the standard Lorenz
attractor is between $d_H = 2.06$ and
$2.16$.\cite{Viswanath04,Martinez93} On the other hand, it is also
known that $d_2 \le d_H$, and this dimension has been estimated to be
$d_2 = 2.05 \pm 0.01$.\cite{GrassbergerPhysicaD} Our estimates are in
line with these results.

\begin{figure*}[htb]
		\subfloat[Full noisy Lorenz
			dynamics]{ \includegraphics[width=0.33\linewidth]{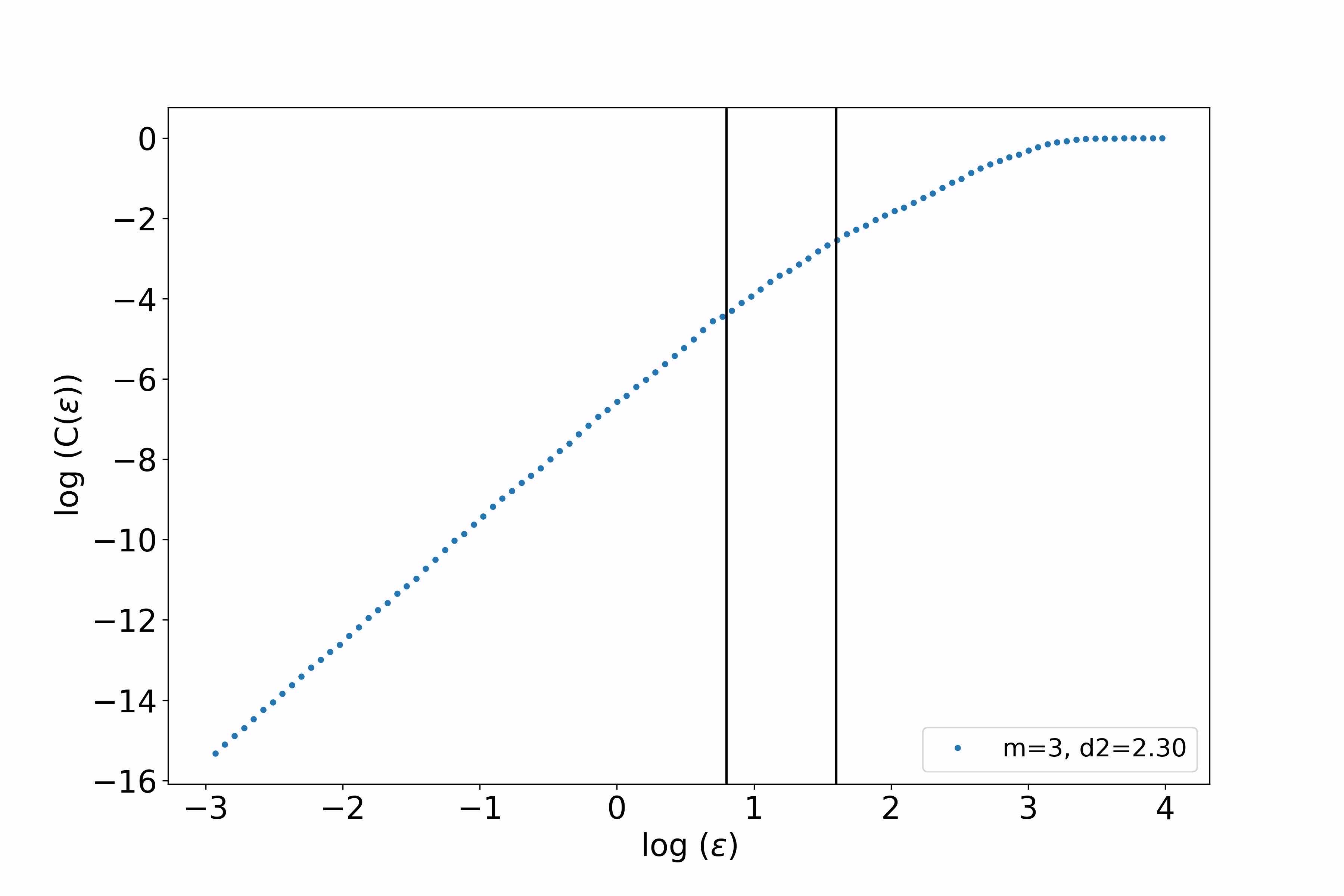}
			} \subfloat[$\tau=21$]{ \includegraphics[width=0.33\linewidth]{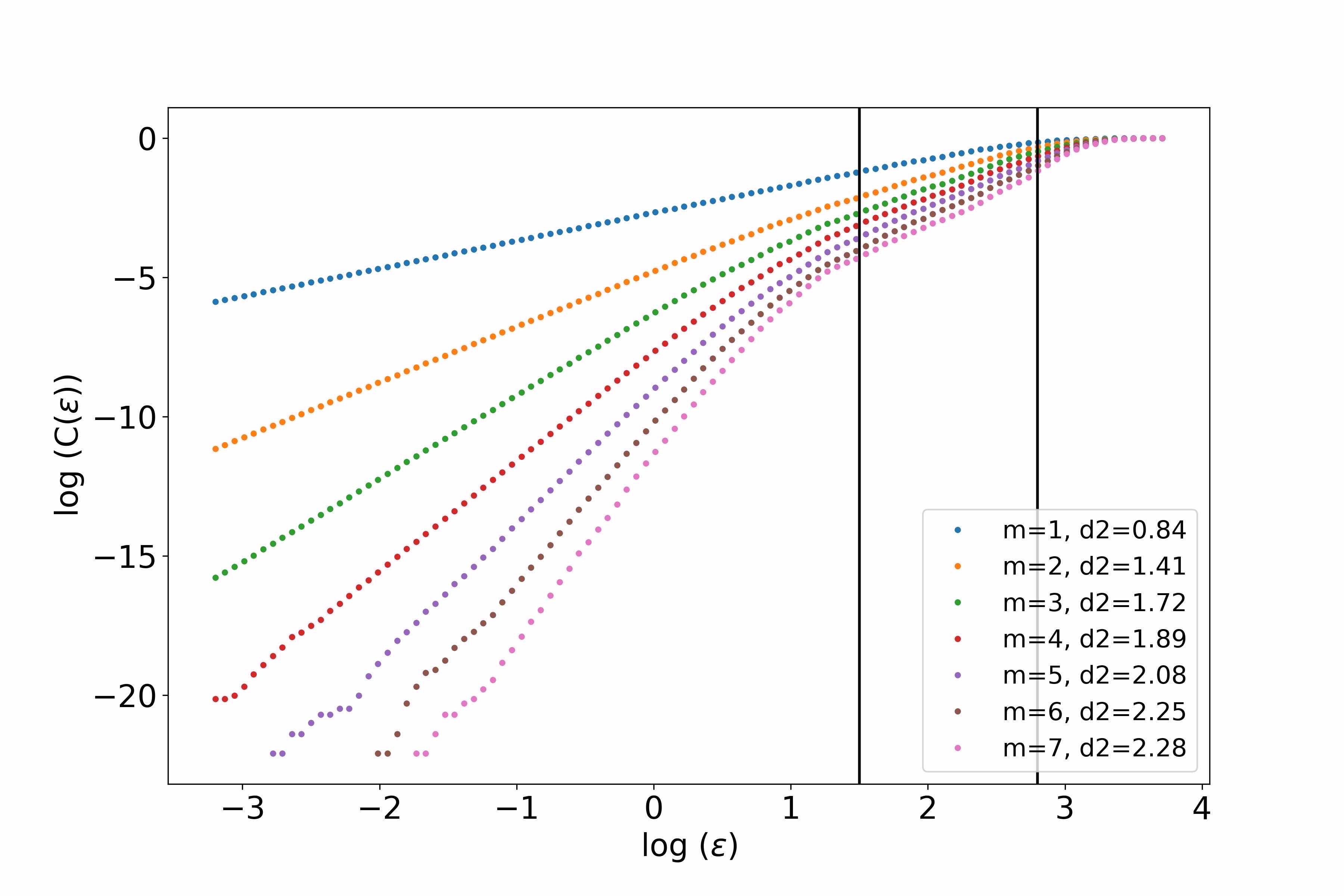}
			} \subfloat[$\tau=60$]{ \includegraphics[width=0.33\linewidth]{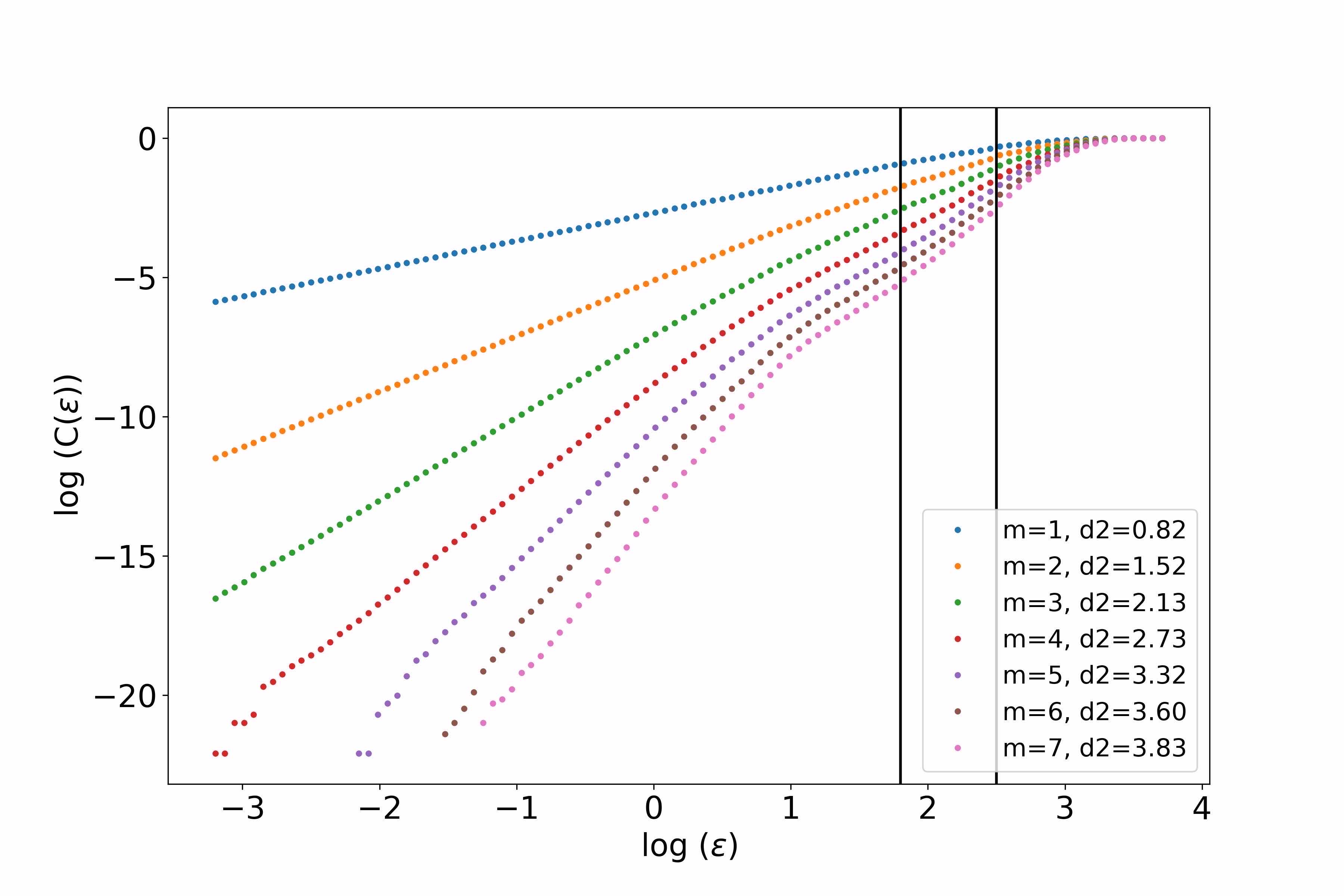}
			} \caption{Log-log plots of $C(\epsilon)$ and
			calculation of the correlation dimension of
			the noisy Lorenz data set for the full
			dynamics, and using $\tau=21$---the
			value suggested by the curvature ---and $\tau=60$, 
			the second, clearer
			minimum of AMI.}  \label{fig:lorenz_noise_d2}
\end{figure*}

For the noisy Lorenz data of \S\ref{sec:noisyLorenz}, the correlation
sum plots for the delay
reconstructions---Fig.~\ref{fig:lorenz_noise_d2}(b) and (c)---have a
more-interesting shape: there is a second scaling region for small
$\epsilon$ that arises from the added noise.\footnote{The Theiler
windows for these calculation, and for the rest of the examples in
this section, were chosen as described above.}  Indeed, when there is
uniformly distributed \textit{iid} noise with a maximum size
$\epsilon_{max}$, then points within a radius $\epsilon_{max}$ will be
dominated by the noise and thus tend to fill out a ball of full
dimension in the embedding space.  Thus for reconstructed data we
expect a computed $d_2 \approx m$ for any scaling domain with
$\epsilon < \epsilon_{max}$.  For Fig.~\ref{fig:lorenz_noise_d2},
$\epsilon_{max} = 1.912$, so $\log(\epsilon_{max}) = 0.65$.  In the
figures, this threshold corresponds to a knee in the curves below
which their slopes are roughly $m$.  This confirms that the lower
scaling region is due to the noise.  The correlation sum plot for $\tC
= 21$ has a relatively broad scaling region with a slope of $d_2 =
2.28$ for $m = 7$, but for $\tAMI = 60$, the scaling region above the
noise threshold is narrower.  This region gives $d_2 = 3.83$, a
significant over-estimate of the correlation dimension of this system.

\begin{figure*}
	\subfloat[$\tau=16$]{
		\includegraphics[width=0.33\linewidth]{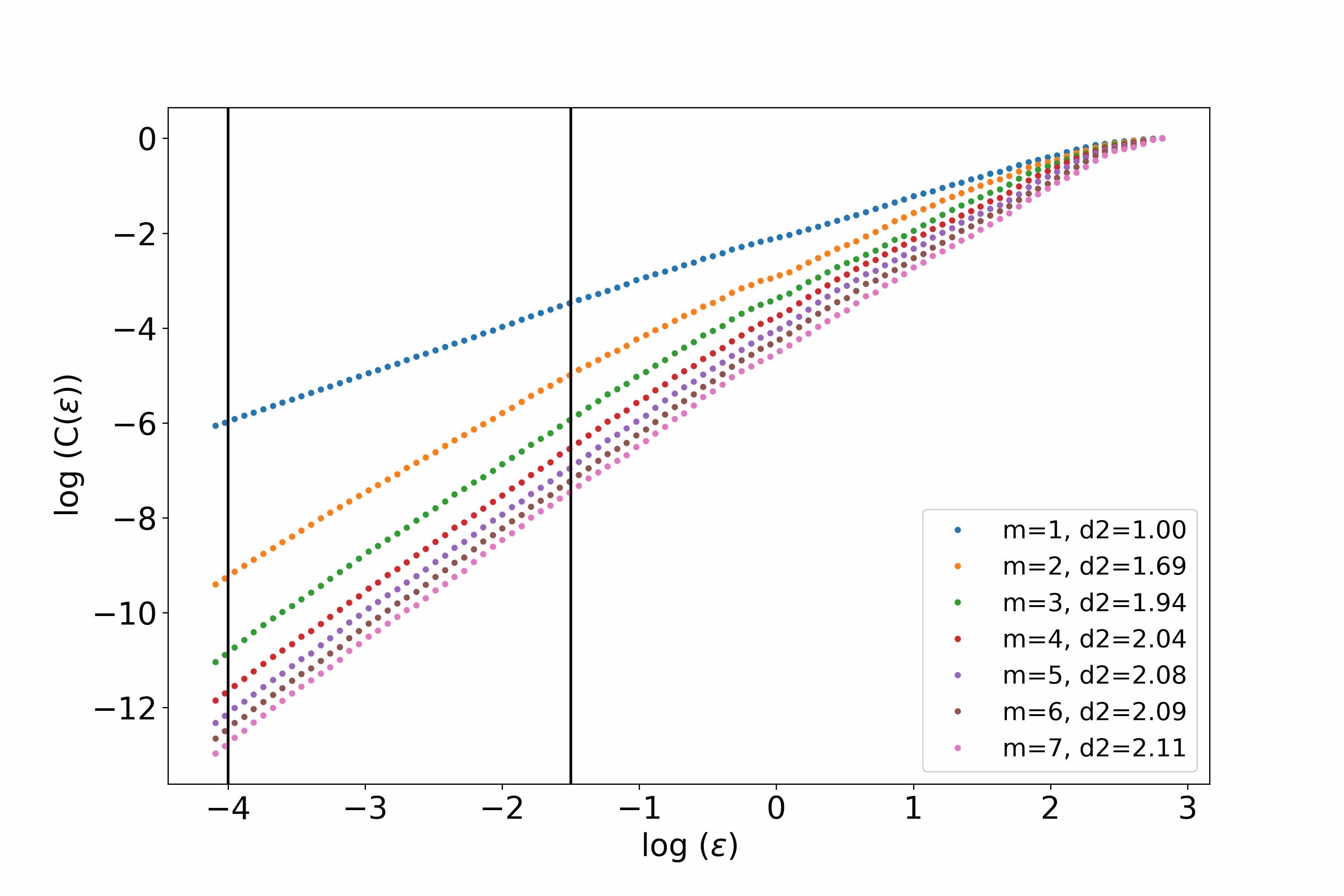}
	}
	\subfloat[$\tau=102$]{
		\includegraphics[width=0.33\linewidth]{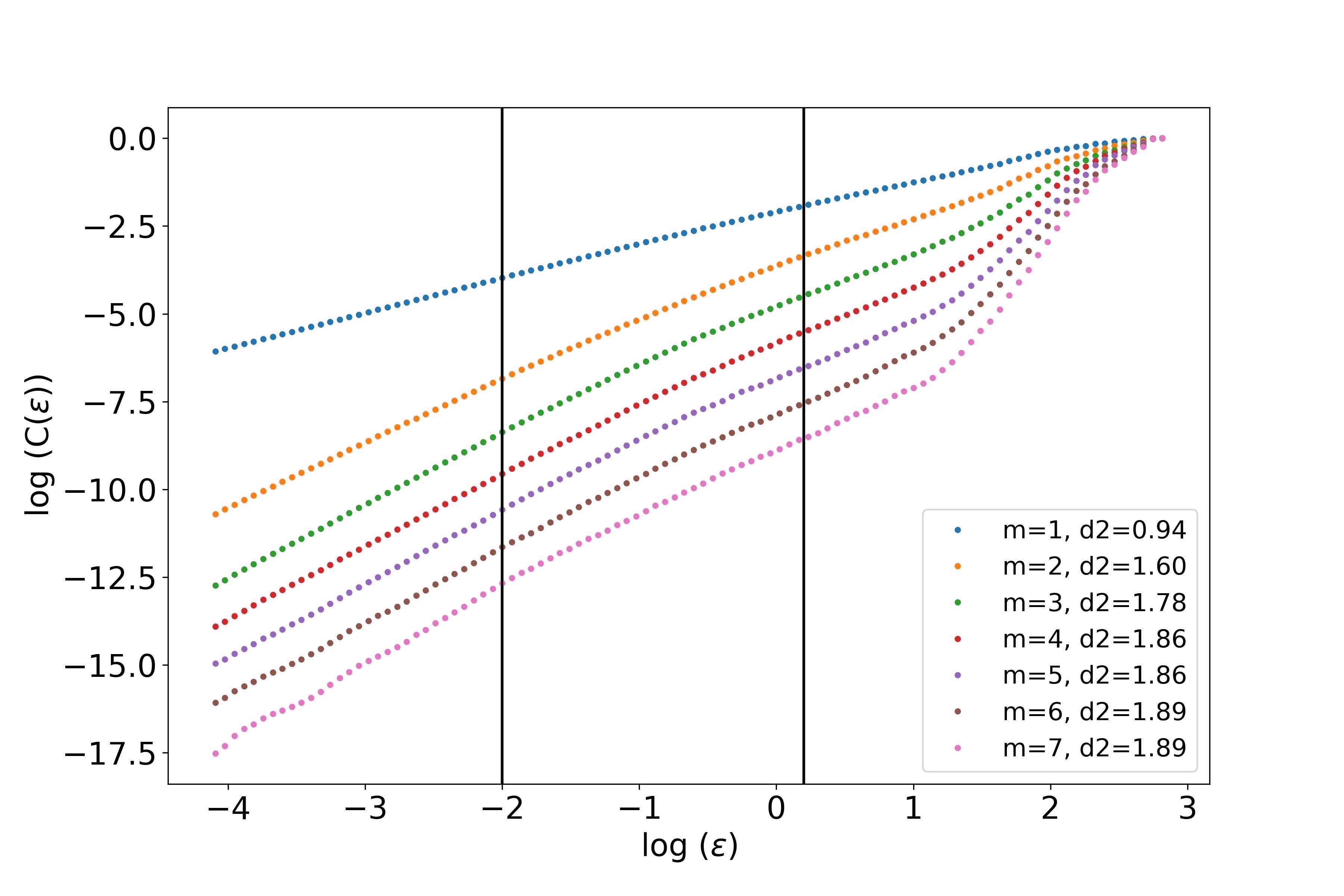}
	}
	\caption{Log-log plots of $C(\epsilon)$ and calculations of
	the correlation dimension of the filtered Lorenz data set
	using $\tau=16$---the value suggested by the curvature
	heuristic--and $\tau=102$, the first clear minimum of AMI. For
	the full state space, from Fig.~\ref{fig:lorenz_d2}, $d_2 =
	2.05$.}  \label{fig:lorenz_filtered_d2}
\end{figure*}

In the third variant of the Lorenz experiments, discussed
in \S~\ref{sec:data-effects}, the goal is to determine whether the
curvature-based heuristic produces a reconstruction that matches the
original dynamics even if the data are distorted by smoothing.  To
explore this, we compared the full dynamics, from
Fig.~\ref{fig:lorenz_d2}(a), to those for  $\tC$ and $\tAMI$
reconstructions of the {\sl filtered} time-series data in
Fig.~\ref{fig:lorenz_filtered_d2}.  The correlation sum plots for the
two reconstructions are significantly different.  The $\tAMI$
embedding in Fig.~\ref{fig:lorenz_filtered_d2}(a) has a
narrow\footnote
{A broader scaling region, {\sl e.g.}, starting at $\log(\epsilon) = -4$, is not
appropriate because of the slight bend in the $m=7$ profile at
$\log(\epsilon) = -2$.}
scaling region, $\log(\epsilon) \in [-2, 0.2]$.
A fit in this region gives the underestimate $d_2=1.89$.
The $\tC$ embedding, on the other hand, has a broad scaling region,
$\log(\epsilon) \in [-4, -1.5]$, yielding $d_2 = 2.11$, closer
to the dimension $2.05$ of the unfiltered, full dynamics.

\begin{figure*}
		\subfloat[Full pendulum dynamics ($10^6$ points)]{
			\includegraphics[width=0.33\linewidth]{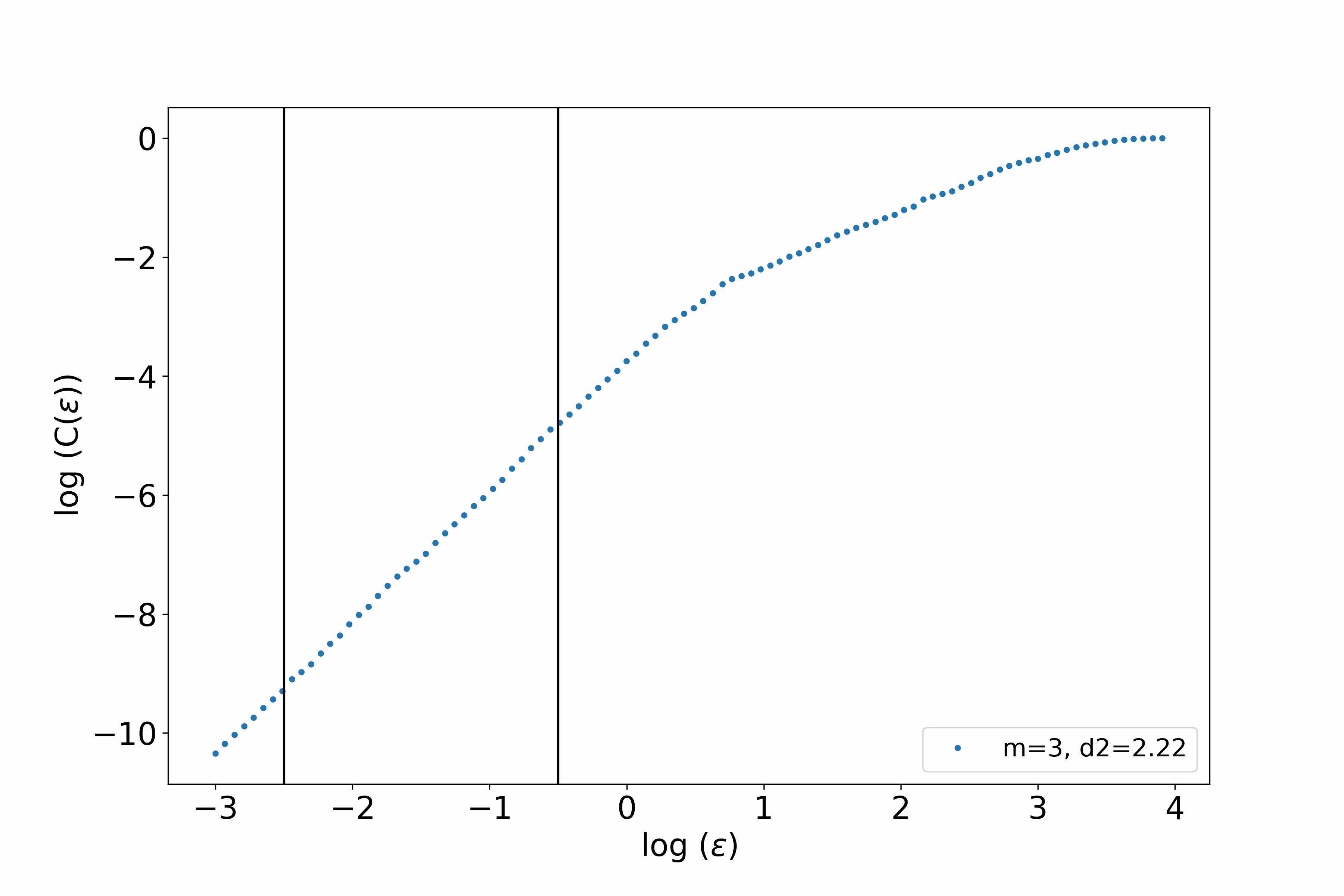}
		}
		\subfloat[$\tau=120$]{
			\includegraphics[width=0.33\linewidth]{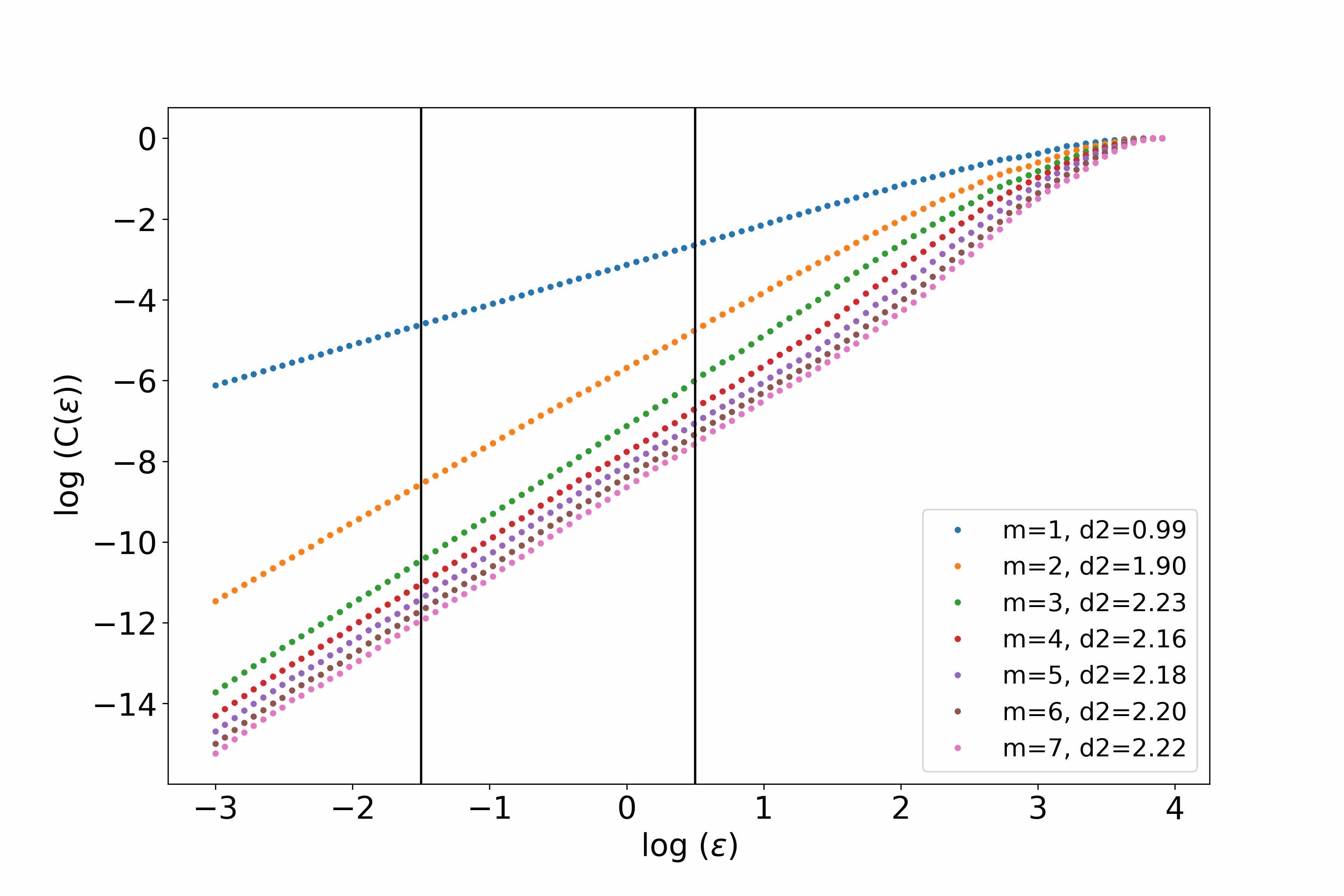}
		}
		\subfloat[$\tau=250$]{
			\includegraphics[width=0.33\linewidth]{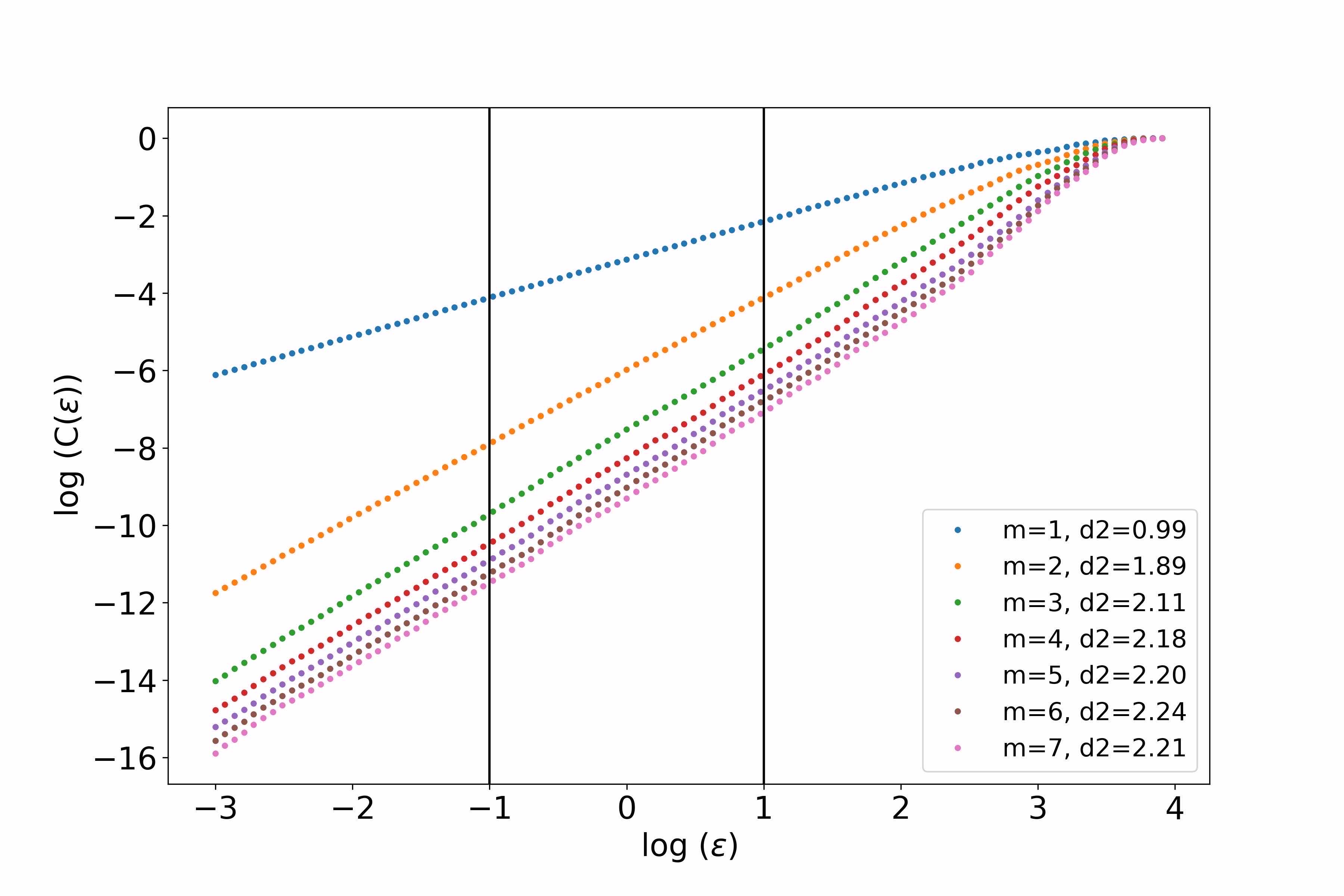}
		}

	\caption{Log-log plots of $C(\epsilon)$ and calculations of
          the correlation dimension of the $10^6$ point pendulum data
          set for the full dynamics giving $d_2 = 2.22$, and using
          $\tau=120$---a representative value in the range suggested
          by the curvature heuristic and $\tau=250$---the value
          suggested by AMI.}  \label{fig:pendulum_d2}

\end{figure*}

\begin{figure*}
		\subfloat[Full pendulum dynamics ($2\times10^5$ points)]{
			\includegraphics[width=0.33\linewidth]{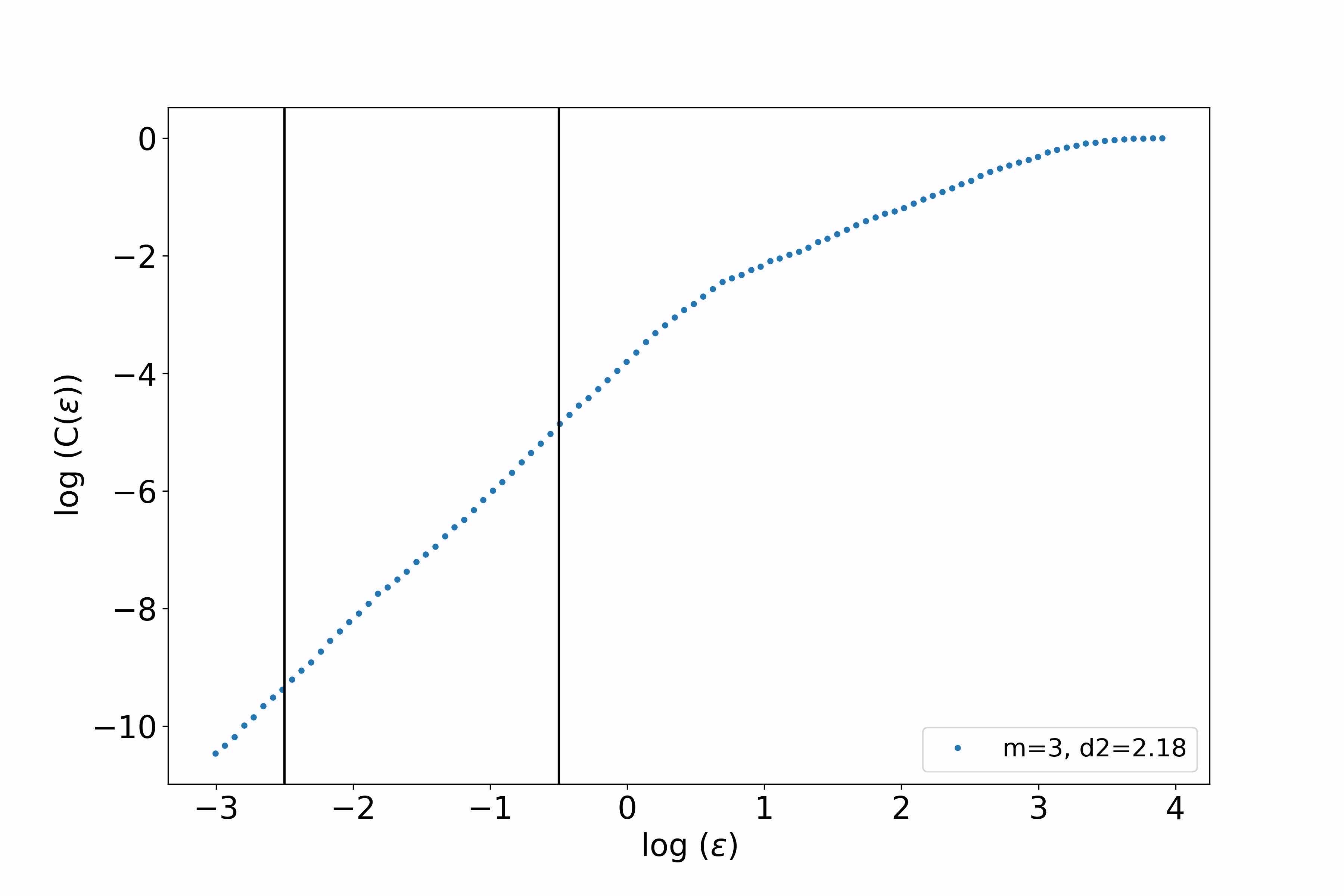}
		}
		\subfloat[$\tau=120$]{
			\includegraphics[width=0.33\linewidth]{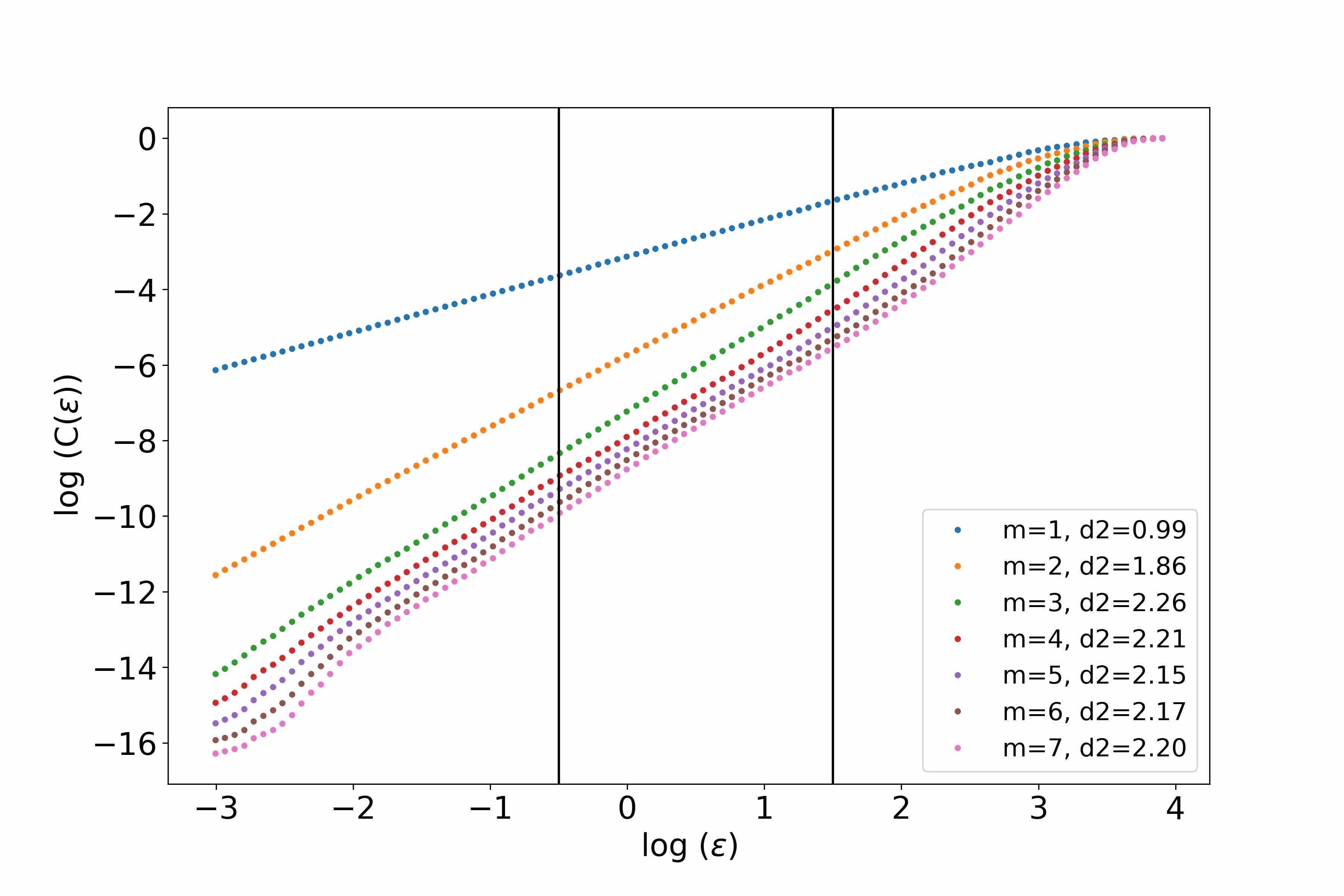}
		}
		\subfloat[$\tau=250$]{
			\includegraphics[width=0.33\linewidth]{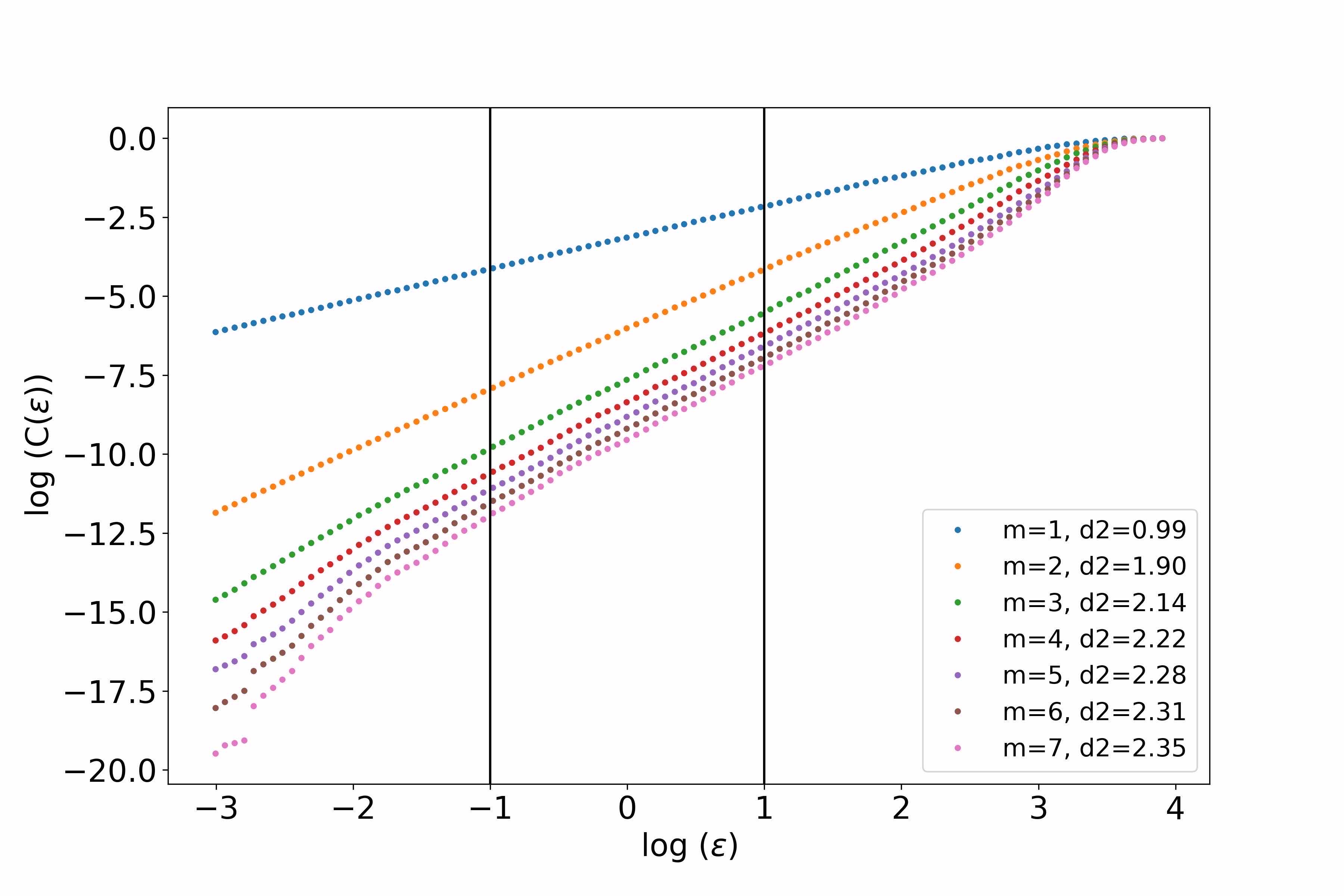}
		}
	\caption{Calculations of the correlation dimension of the
          shorter pendulum data set for the full dynamics giving $d_2
          = 2.18$, and using $\tau=120$------a representative value in
          the range suggested by the curvature heuristic---and
          $\tau=250$, the value suggested by
          AMI.}  \label{fig:pendulum_200K_d2}
\end{figure*}

Figs.~\ref{fig:pendulum_d2} and~\ref{fig:pendulum_200K_d2} show the
correlation plots for the driven damped pendulum example discussed in
\S\ref{sec:Pendulum} and \S\ref{sec:data-effects}, respectively.
For the $10^6$-point trajectory, the correlation plot for the full
dynamics, Fig.~\ref{fig:pendulum_d2}(a), has a broad scaling region
$\log(\epsilon) \in [-2.5, 0.5]$; a fit for this region yields $d_2 =
2.22$.  The $\tC$ and $\tAMI$ reconstructions of these data in
panels (b) and (c) of the figure also have broad scaling regions
($\log(\epsilon) \in [-1,1.5]$), with $d_2 = 2.22$ and $d_2 = 2.21$,
respectively---both good approximations to the true value.  The
scaling regions for the shorter pendulum data set, shown in
Fig.~\ref{fig:pendulum_200K_d2}, are identical to those for the longer
trajectory, but their slopes are different.  For the $\tau_I = 120$
reconstruction, $d_2 = 2.20$, closely matching that for the full dynamics, $d_2
= 2.18$. The $\tAMI = 250$ reconstruction, on the other hand,
overestimates the dimension, giving $d_2 = 2.35$.

\end{document}